\documentclass[10pt,twocolumn,preprintnumbers,amsmath,amssymb,nofootinbib
,superscriptaddress]{revtex4-1}
\usepackage{graphicx,longtable,mathrsfs,color,array}
\usepackage{hyperref}
\usepackage[usenames,dvipsnames]{xcolor} 
\usepackage{amssymb,amsmath,mathtools,mathrsfs,slashed} 
\usepackage{epsfig,subfigure,placeins,float} 
\usepackage{booktabs,longtable,ctable,multirow} 
\usepackage{exscale,relsize} 
\usepackage[normalem]{ulem} 
\usepackage{enumerate}
\usepackage{enumitem}
\usepackage{comment}
\usepackage{color}
\usepackage{braket}
\allowdisplaybreaks[1]

\newcommand{\be}{\begin{equation}}
\newcommand{\ee}{\end{equation}}
\newcommand{\bea}{\begin{eqnarray}}
\newcommand{\eea}{\end{eqnarray}}

\begin{document}
\title{Cosmology of Gravi-Axions}

\author{Stephon Alexander}
\email{stephon\_alexander@brown.edu}
\affiliation{Brown Theoretical Physics Center and Department of Physics, Brown University, Providence, Rhode Island 02903, United States}
\author{Gregory Gabadadze}
\email{gg32@nyu.edu}
\affiliation{Center for Cosmology and Particle Physics, Department of Physics, New York University, New York, New York 10003, United States}
\author{Leah Jenks}
\email{ljenks3@jh.edu}
\affiliation{Kavli Institute for Cosmological Physics, University of Chicago,  Chicago, IL 60637 USA}
\affiliation{William H. Miller III Department of Physics \& Astronomy,\\ Johns Hopkins University, 3400 N. Charles St., Baltimore, MD 21218, USA}
\author{Nicol\'as Yunes}
\email{nyunes@illinois.edu}
\affiliation{Illinois Center for Advanced Studies of the Universe, Department of Physics, University of Illinois at Urbana-Champaign, Urbana, Illinois 61801, United States}
\date{Received \today; published -- 00, 0000}

\begin{abstract}

We show how a mass term for gravitational axions (``gravi-axions'') with a Chern-Simons coupling to gravity can naturally arise due to non-perturbative contributions from Euclidean wormholes, breaking the continuous shift symmetry of the standard theory. The induced mass can be generated in a cosmologically relevant range to be the dark matter or dark energy of the universe for a reasonable and well-motivated range of the symmetry breaking scale. Upon generating the gravi-axion mass term, we discuss the cosmology of the theory. We find that the gravi-axion can be produced to be the dominant dark matter component via misalignment or gravitational particle production, and that in a different regime, the gravi-axion can act as dynamical dark energy. We discuss gravi-axion decay into gravitons as a potential observational window for this theory.
Finally, we find that, for late-time,  astrophysical compact objects, cosmologically-relevant gravi-axions behave as minimally-coupled, massive scalar fields.

\end{abstract}

\date{\today}
\maketitle


\section{Introduction}

Axions are ubiquitous in particle theory and models of cosmology. These pseudo-scalar fields arise in a variety of different theoretical contexts. The Peccei-Quinn (PQ) symmetry was introduced to solve the 
strong CP problem \cite{Peccei:1977hh,Peccei:1977ur}, and the 
axion was identified as a low-energy remnant of the PQ symmetry 
solving the problem in QCD \cite{Weinberg:1977ma,Wilczek:1977pj,Kim:1979if,
Shifman:1979if,Zhitnitsky:1980tq,Dine:1981rt}. It is characterized by a coupling to the gluon (photon) field strength and its dual. Another class of axions (often referred to as `axion-like particles' or `ALPs') that couple to gauge fields are predicted from string theory \cite{Svrcek:2006yi, Masso:1995tw, Masso:2002ip, Conlon:2006tq}, leading to the so-called `string axiverse' \cite{Arvanitaki:2009fg}. Axions have also been studied  intensively in a phenomenological context (see the reviews \cite{Sikivie:2006ni, Marsh:2015xka} and references therein). In all cases, axions are of significant theoretical interest for beyond-the Standard Model physics, including as a candidate for dark matter (e.g. \cite{Hu:2000ke, Hui:2021tkt}) or dark energy (e.g. \cite{Frieman:1995pm,Carroll:1998zi}).

Less well studied in these contexts is the gravitational axion, or `gravi-axion,' which couples directly to the Riemann tensor and its dual, also known as the Pontryagin density. This type of non-minimal coupling is well known in the context of dynamical Chern-Simons (dCS) gravity, an extension of general relativity (GR) that is characterized by the inclusion of a pseudoscalar field coupled to the Pontryagin density \cite{Lue:1998mq, jackiw, Alexander:2009tp}. dCS gravity has both particle physics \cite{ALVAREZGAUME1984269} and string theory \cite{Polchinski:1998rr, PhysRevLett.96.081301, Green:1987mn, Alexander:2004xd} motivations, and can also be generated from a variety of progenitor theories \cite{Alexander:2021ssr, Alexander:2022cow}. This theory has been well studied in many contexts, including for black holes \cite{Grumiller:2007rv, Shiromizu:2013pna, Yunes:2009hc,Cardoso:2009pk,  Yagi:2012ya, Alexander:2022avt}, gravitational waves \cite{sopuertayunes, Alexander:2007kv, Yunes:2010yf, Alexander:2017jmt, jenks:2023pmk, Lagos:2024boe,Daniel:2024lev,Callister:2023tws}, and the early universe \cite{Lue:1998mq, Alexander:lepto, alexander:2011hz, Inomata:2024ald}, since it can leave distinct imprints on cosmological and gravitational observables (see \cite{Alexander:2009tp} and references therein).

As mentioned above, the implications for the gravi-axion on dark matter, dark energy, or other cosmological relics are less well studied. It would be cosmologically convenient for the gravi-axion to comprise a component of the dark matter or the dark energy, in that their presence could be elegantly explained by an additional field arising from the gravitational sector, requiring no \textit{a priori} coupling to Standard Model (SM) particles. Such a scenario has not been seriously considered previously because, in the canonical instantiation of dCS gravity, the corresponding pseudo-scalar is taken to be massless. This is due to the presence of a continuous shift symmetry in the theory, which is thought to protect the pseudo-scalar from generating a potential. Some previous studies have considered the effects of a massive dCS field by adding a mass `by hand' \cite{Smith:2007jm, Nashed:2022kes, Richards:2023xsr}, but without a first principles explanation for the breaking of the shift symmetry. Therefore, the dCS axion is generally treated as a massless background field that becomes relevant in regimes of large curvature.

Even if a classical theory contains a massless scalar protected by a global symmetry (such as the shift symmetry), however, one expects that non-perturbative quantum effects will generate a mass and potential. For example, the QCD axion acquires its mass due to non-perturbative effects of gluons, and non-perturbative gravitational effects are also well-known to generate a contribution to the potential leading to global symmetry breaking \cite{Lee:1988ge, Abbott:1989jw, Rey:1989mg,Grinstein:1988wr,Brown:1989df, Kallosh:1995hi}. This is a result of the usual adage ``no global symmetries in quantum gravity,'' an old idea that goes back to the study of Euclidean wormholes, or ``baby universes,'' in the 1980s, e.g., \cite{Giddings:1987cg, Coleman:1988tj, Preskill:1988na, Klebanov:1988eh,Choi:1989ck,Preskill:1989zu}. In this scenario, the high-energy effects of a theory consisting of many baby Euclidean universes can be described by adding a correction to the low-energy effective action of the theory. Recent work, in fact, has further explored the possibility for generating the mass of an axion dark matter candidate this way \cite{Alonso:2017avz, Hebecker:2018ofv, Cheong:2024kum,Cheong:2023hrj}. 

In this work, we show how the effects of Euclidean wormholes endow the gravi-axion with a mass. Beginning from a weakly coupled ultraviolet (UV) progenitor theory, we show that the low-energy effective field theory includes a dCS coupling for the gravi-axion, as well as a mass term generated by Euclidean wormhole effects. This yields a massive gravi-axion theory that can have cosmological effects. We consider two concrete wormhole solutions, the Giddings-Strominger wormhole \cite{Giddings:1987cg}, and the string-inspired wormhole discussed in \cite{Kallosh:1995hi}. In the former case, for values of the PQ scale of the theory, $\mu$, near $\mu\sim 10^{16} - 10^{17}$ GeV, we find that the gravi-axion can be generated in a mass range that is cosmologically relevant for dark matter or dark energy, but is exponentially sensitive to the coupling. In the latter scenario, we find that the gravi-axion can have masses in the ultralight regime for a wider range of $\mu$. In the dark-matter regime, the gravi-axion can be produced via misalignment or gravitational particle production. In the dark-energy regime, the gravi-axion acts as a quintessence field and may provide intriguing connections to recent observational results hinting at evolving dark energy \cite{DESI:2024mwx, DESI:2025zgx}.

In principle, this scenario also yields a theory of massive dCS gravity. However, in our semi-classical calculation, we find two possibilities. One possibility arises if we choose the PQ scale so that the non-minimal coupling between the gravi-axion and the Pontryagin invariant is non-negligible; in this case, the gravi-axion mass is exponentially suppressed and the theory reduces to a massless dCS gravity theory, removing any motivation to study astrophysical observational signatures of massive dCS fields.
The other possibility arises if we choose the PQ scale to generate a non-negligible mass for the gravi-axion; in this case, the non-minimal coupling between the axion and gravity (through the Pontryagin invariant) is highly suppressed at late times, and the theory reduces effectively to that of a massive axion field that couples minimally to gravity. However, due to large field values in the early universe, primordial effects may imprint themselves in late-time observables. We study one such example, by considering the gravitational-wave (stochastic background) signature generated from primordial axion decay to gravitons. We find that this decay produces a characteristic high-frequency gravitational-wave background when the gravi-axions are supermassive. 

The structure of the paper is as follows.  In Sec.~\ref{sec:dcswormholes}, we show how the gravi-axion coupling term and mass can be generated via Euclidean wormhole effects, beginning from a weakly-coupled  progenitor field theory. In Sec.~\ref{sec:DM}, we discuss the phenomenology of the gravi-axion as a dark matter candidate. We discuss the gravi-axion decay into gravitons as a potential observable signature of this scenario in Sec.~\ref{sec:decay}.  We show that the gravi-axion can act as a quintessence field in Sec.~\ref{sec:DE}, and finally, we conclude with a discussion and comment on future directions in Sec.~\ref{sec:conclusions}.  In Appendices~\ref{ssec:instantons} and ~\ref{ssec:GS}, we provide further details on the Euclidean wormhole procedure to generate the gravi-axion potential and on the Giddings-Strominger wormhole solution, respectively. Throughout the paper, we work in natural units such that $c = 1 = \hbar$ unless otherwise specified, and use a mostly minus metric signature.

\section{Gravitational Axions from Euclidean Wormholes}
\label{sec:dcswormholes}
In this section, we discuss how massive gravi-axions are generated from a weakly-couled theory containing chiral fermions and a complex scalar field. This procedure is analogous to the generation of a potential contribution for the QCD or ultralight axion, as reviewed in e.g. \cite{Alonso:2017avz, Hebecker:2018ofv}. Further details on the standard mechanism can be found in Appendix~\ref{ssec:instantons}, while here we show how this mechanism is applied in the gravitational scenario to our theory.

Gravi-axions can be thought of as particles that arise from an effective field theory (EFT), after integrating out massive degrees of freedom in some other, weakly coupled UV theory. References \cite{Alexander:2021ssr,Alexander:2022cow} showed that one viable progenitor theory is one that contains chiral fermions with a Yukawa coupling to a complex scalar field with a quartic potential. The Lagrangian for the weakly-coupled UV theory is given by 
\begin{align}
    \mathcal{L}_{\rm UV} &= -\kappa R + |\nabla\Phi|^2 + \lambda_1\left(|\Phi|^2 - \frac{\mu^2}{2}\right)^2\nonumber \\
    &+ \bar{\Psi}_Li\slashed{\nabla}\Psi_L + \bar{\Psi}_Ri\slashed{\nabla}\Psi_R - \lambda_2(\bar{\Psi}_L\Phi \Psi_R + \rm{h.c.}), 
    \label{eq:UVaction}
\end{align}

where $\kappa =(16\pi G)^{-1}$, $R$ is the Ricci scalar, $\Phi$ is a complex scalar field that may be decomposed via
\be 
\Phi = \frac{\mu}{{\sqrt{2}}} \; e^{i\varphi/\mu}, 
\label{eq:phidecomp}
\ee 
with vacuum expectation value (VEV) $\mu = \sqrt{2}\langle \Phi \rangle$. The chiral fermions are $\Psi_{R,L}$, and $\slashed{\nabla} = \gamma^\mu \nabla_\mu$, where $\gamma^\mu$ are the gamma matrices and $\nabla_\mu$ is the covariant derivative with spacetime indices, $\mu$. Finally, $\lambda_{1,2}$ are coupling constants, and h.~c.~ stands for Hermitian conjugate. 

Given the above weakly coupled theory, the gravi-axion EFT is generated by the gravitational axial anomaly \cite{Delbourgo:1972xb} as follows. The theory defined by Eq.~\eqref{eq:UVaction} is symmetric under global U(1) axial Peccei-Quinn transformations. This symmetry is spontaneously broken due to the nonzero VEV of $\varphi$, generating masses for the fermions and for the modulus of the complex scalar. These massive modes can be integrated out from the action. Then at low energies, the anomalous axial diagrams give rise to 
\be 
\mathcal{L}_\varphi^{\rm CS} =\frac{\varphi}{4\mu} R \tilde{R}, 
\label{eq:LCS}
\ee 
and the EFT Lagrangian density becomes
\be
{\cal{L}}_{\rm EFT} = -\kappa R +\frac{1}{2}(\nabla_\mu \varphi) (\nabla^\mu \varphi) + \frac{\varphi}{4\mu} R \tilde{R}+...\,,
\label{eq:EFT}
\ee

which couples the Pontryagin invariant $R\tilde{R}$, defined in terms of the Riemann tensor and its dual via 
\be 
R\tilde{R} = \frac{1}{2}\epsilon^{\rho\sigma\alpha\beta}R^\mu{}_{\nu\alpha\beta}R^\nu{}_{\mu\rho\sigma},
\label{eq:pont}
\ee
to the pseudoscalar, $\varphi$~\cite{Alexander:2021ssr}. The VEV $\mu$ now acts as an energy scale that controls the magnitude of the non-minimal coupling between the pseudo-scalar and gravity. 

The EFT 
\eqref{eq:EFT} should be viewed as truncation of an infinite series of higher dimensional terms suppressed by powers of the scale $\mu$. 
Hence, in the regime of applicability of \eqref{eq:EFT} 
the last term in ${\cal{L}}_{\rm EFT}$ can only be viewed as a small correction to the first two terms, but it is the leading correction 
among the infinite number of the higher dimensional terms. That said, 
in the present example, all of the higher dimensional terms are calculable within the progenitor weakly coupled UV theory \eqref{eq:UVaction}.

 The action defined by Eq.~\eqref{eq:EFT} has shift symmetry under $\varphi \rightarrow \varphi + const.$, for topologically trivial space-times; this prevents $\varphi$ from generating a mass perturbatively. However, as was pointed out in \cite{Alexander:2021ssr}, one expects that non-perturbative quantum gravity effects will break the shift symmetry and generate a potential. This potential has an exponential suppression, which is therefore why it is generally assumed to be negligibly small. However, in this work, we will characterize this contribution and show that it can have non-trivial consequences in 
 early-time cosmology. 

The physics responsible for explicitly breaking the shift symmetry and generating a mass term for the gravi-axion is the effect of Euclidean wormholes at high energies. Wormholes represent non-perturbative fluctuations in the spacetime topology. Integrating over these wormhole configurations in the gravitational path integral induces additional terms in the low-energy effective action which are exponentially dependent on wormhole action, $S_W$. In the case of the theory of a massive, complex scalar field $\Phi$, these terms provide a potential, $V$ for the field of the form $V \sim \Phi^n + {\rm h.c.}$, with coefficients controlled by the wormhole action. 
The potential explicitly breaks the U(1) symmetry and, inserting Eq.~\eqref{eq:phidecomp}, it generates an axion-like cosine potential. Such a mechanism has been studied in detail in the past~\cite{Lee:1988ge, Abbott:1989jw, Rey:1989mg,Grinstein:1988wr,Brown:1989df, Kallosh:1995hi} for complex scalar field theories, without considering the additional coupling to fermions, and is the gravitational analog to the generation of a potential via instantons in QCD \cite{Schafer:1996wv}.  

Let us then begin by considering the correction to the theory defined by Eq.~\eqref{eq:UVaction} that arises from integrating over the Euclidean wormhole configurations in the gravitational path integral. We will here highlight the main relevant points of the path integral calculation, delegating further details to Appendix~\ref{ssec:instantons}, which can also be found readily in the original literature \cite{Lee:1988ge, Abbott:1989jw, Rey:1989mg,Grinstein:1988wr,Brown:1989df, Kallosh:1995hi} and reviews \cite{Alonso:2017avz, Hebecker:2018ofv}. We expect that the wormhole contributions will be relevant at high energies, near the Planck scale $M_P$. At the same time,  the PQ scale of our effective theory, $\mu$, needs to satisfy $\mu \ll M_{P}$ for it to be tractable. From our Lagrangian density, Eq.~\eqref{eq:UVaction}, note that the additional coupling to fermions will not alter the topological contributions from the wormholes. This is because the Yukawa couplings generate fermion masses once $\Phi$ acquires a VEV, but do not change the zero modes in the gravitational and scalar sector, which are integrated over to obtain the wormhole induced correction. From the wormhole contributions, the relevant term to add to the effective Lagrangian is \cite{Abbott:1989jw} (see also further details in Appendix~\ref{ssec:instantons}):
\be 
\mathcal{L}_{\rm eff} \supset \sum_{n=1}^{\infty}\alpha_n g_n \Phi^n + \rm{h.c.}, 
\label{eq:LWH}
\ee 
where $\alpha_n$ is a constant that is in general difficult to compute. The prefactor $g_n \propto e^{-S_W}$ reflects the exponential suppression by the wormhole action, $S_W$. The Lagrangian of Eq.~\eqref{eq:UVaction} then becomes: 
\begin{align}
    \mathcal{L}_{\rm eff} &= - \kappa R + |\nabla\Phi|^2 + \lambda_1\left(|\Phi|^2 - \mu^2\right)^2\nonumber \\
    &+ \bar{\Psi}_Li\slashed{\nabla}\Psi_L + \bar{\Psi}_Ri\slashed{\nabla}\Psi_R - \lambda_2(\bar{\Psi}_L\Phi \Psi_R + \rm{h.c.}) \nonumber \\
    &+\sum_{n=1}^{\infty}\left(\alpha_n g_n\Phi^n + \rm{h.c.}\right). 
\end{align}

Having determined the wormhole correction, we can now obtain the usual gravi-axion dCS term by calculating the relevant
fermion loop diagrams. Doing so, leaves only the massless phase, $\varphi$, and the dCS term arises from the gravitational axial anomaly triangle diagram, as discussed in \cite{Alexander:2021ssr}. One may worry that the addition of the new term from the wormholes will alter the dCS correction; however, the anomalous part of the triangle fermion loop is unaffected and we can proceed as usual. The effective action then becomes

\begin{align} 
\mathcal{L}_{\rm eff} &=- \kappa R + \frac{1}{2}\nabla_{\mu}\varphi \nabla^{\mu}\varphi + \frac{\varphi}{4\mu}R\tilde{R}\nonumber\\
&+\sum_n  \frac{\alpha_n\mu^n g_n}{2^{n-1}}\cos\left(\frac{n\varphi}{\mu}\right),
\label{eq:Seff1}
\end{align}

where we can see that the continuous shift symmetry of the theory has been broken to the discrete $2\pi$ periodicity,  by the addition of the potential term. 
One may be concerned that we must include higher-order derivative terms in the wormhole correction, e.g.~a term such as $(\partial \varphi)^n$. However, each derivative will induce an additional suppression of $\mu$, and thus, for every derivative term, there will be a non-derivative term that dominates over it. We can therefore indeed just keep the contribution from Eq.~\eqref{eq:LWH}. 

In the semiclassical approximation, the wormhole contribution will be dominated by those with the smallest action, so we will consider only the $n=1$ contribution. On dimensional grounds, one can assume that $\alpha_1g_1\mu \sim \mathcal{O}(L^{-4})$, where $L$ is the length of the wormhole throat, such that we can approximate the potential as 
\be 
V_{\varphi} \approx L^{-4}e^{-S_W}\cos\left(\frac{\varphi}{\mu}\right).
\ee 

To explicitly see the mass term, we can expand the cosine in small $\varphi/\mu$ to obtain
\begin{align} 
\mathcal{L}_{\rm eff} = - \kappa R & + \frac{1}{2}\nabla_a\varphi \nabla^a\varphi + \frac{\varphi}{4\mu}R\tilde{R} - \frac{1}{2}\frac{e^{-S_W}}{L^4\mu^2}\varphi^2,
\label{eq:Seff-final}
\end{align}
where we have dropped the constant term that does not contribute to the field equations. We then have an explicit mass term for $\varphi$ of the form, 
\begin{equation}
m_{\varphi} = \frac{e^{-S_W/2}}{L^2 \mu}.
\end{equation}

Let us now discuss what mass range the gravi-axion can be produced in to be cosmologically relevant. The exponential suppression by the wormhole action makes the mass extremely sensitive to the PQ scale, but will also allow for a wide range of masses to be produced. Thus far, we have not specified the wormhole action in order to illustrate the most generic physics. However, to discuss the cosmological consequences of the theory, we will consider two concrete examples. First, we will specify to the Giddings-Strominger (GS) wormhole \cite{Giddings:1987cg}\footnote{Note that the usual Giddings-Strominger wormhole solution comes from a complex scalar field theory that does not include the dCS term. However, since the Pontryagin density $\tilde{R}R$ vanishes for spherically symmetric spacetimes, the presence of the $\varphi R\tilde{R}$ term does not modify the Giddings-Strominger solution. More general wormhole geometries with $\tilde{R}R \neq 0$ could lead to further corrections, but here we just consider the simplest case.}. In this case, the wormhole throat length, $L$, is given by 
\be 
L^{\rm GS} = \left(\frac{1}{3\pi^3 M_P^2 \mu^2}\right)^{1/4},
\ee 

and the full wormhole action including boundary terms is
\be 
S_W^{\rm GS} = \frac{\sqrt{3\pi}}{8}\frac{M_P}{\mu}\left(1 - \frac{2}{\pi}\right).
\ee 
As long as $\mu<< M_P$, the euclidean action, $S_W^{\rm GS}>>1$, and 
the quasi-classical approximation is justified. 
This leads to a corresponding gravi-axion mass that is characterized by the Planck mass and the PQ scale, $\mu$: 

\be 
m_\varphi^{\rm GS} \approx \sqrt{3\pi^3}M_P e^{-\frac{\sqrt{3\pi}}{16}\frac{M_p}{\mu}\left(1 - \frac{2}{\pi}\right)}.
\label{eq:mGS}
\ee 

\begin{figure}[htb]
    \centering
    \includegraphics[width=0.5\textwidth]{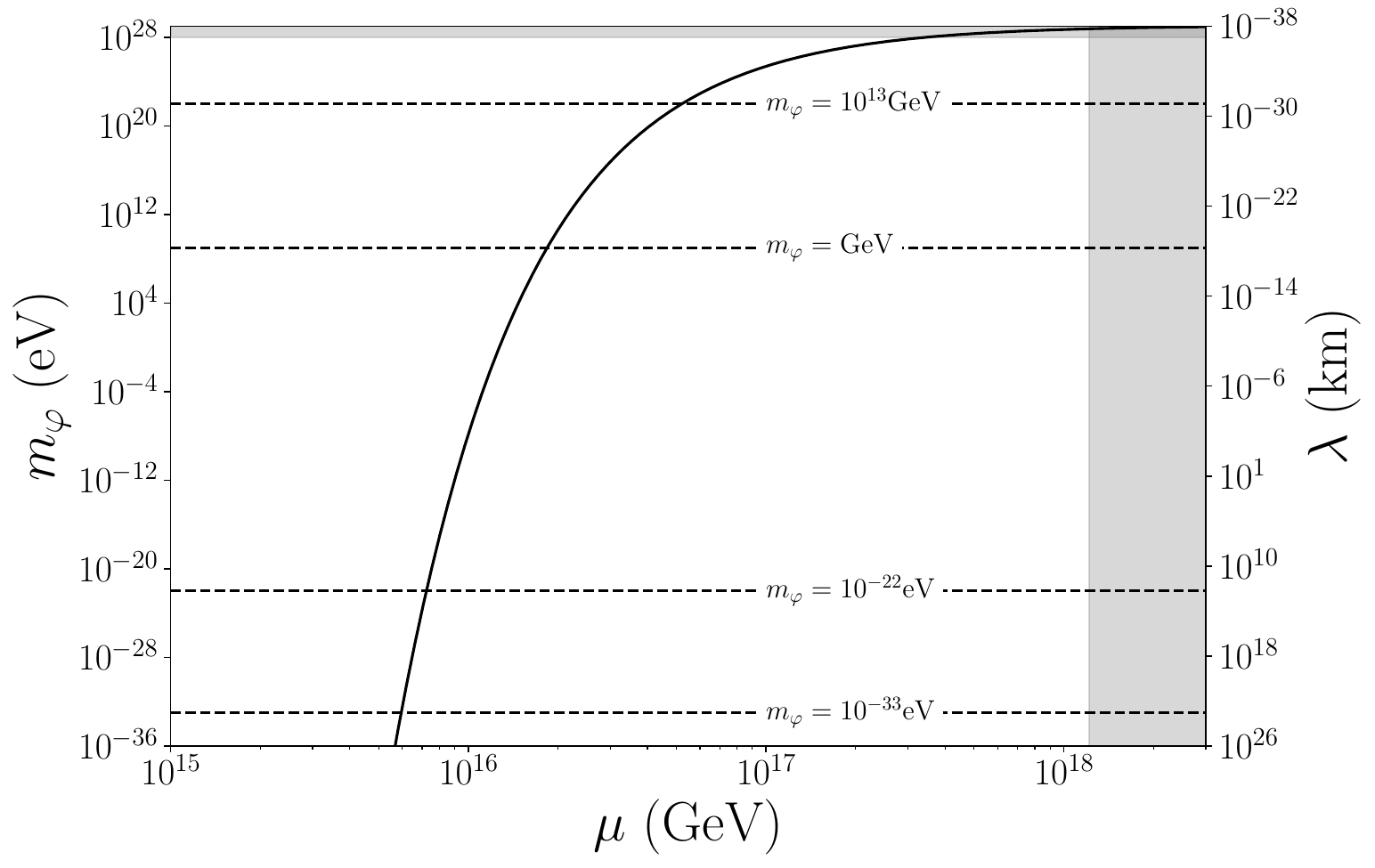}
    \caption{
    Mass of the gravi-axion in terms of the PQ scale, $\mu$, assuming the Giddings-Strominger wormhole solution. We denote several benchmark mass values for cosmologically relevant quantities. The horizontal gray shaded region indicates where $m_\varphi > M_P$, while the vertical gray shaded region denotes where the EFT condition, $\mu \ll M_p$ is no longer valid. We denote the compton wavelength, $\lambda$, of the gravi-axion on the right y-axis.
    }
    \label{fig:n1mass}
\end{figure}

Figure~\ref{fig:n1mass} shows the gravi-axion mass in eV for a range of $\mu$. Recall that, for the EFT to remain valid, we must have $\mu \ll M_P$, and we here ensure this condition is satisfied by requiring that $\mu < 0.1 M_P$ strictly. 
The region where this condition is violated is denoted by the vertical gray shaded region. For $\mu$ close to the maximum of the allowed region, we obtain gravi-axions with super-Planckian masses, denoted by the horizontal, gray-shaded region. For the range of $\mu$ that allows us to remain within the regime of validity of the EFT and generate sub-Planckain masses, $\mathcal{O}(10^{15}) < \frac{\mu}{\rm GeV} < \mathcal{O}(10^{17})$, we find that we can obtain gravi-axions with extremely heavy masses, e.g. $m_\varphi > 10^{13}$ GeV, which become exponentially smaller as one lowers $\mu$ to find $m_\varphi < 10^{-33}$ eV. Despite this exponential sensitivity, we see that it is indeed possible to obtain a gravi-axion with a cosmologically relevant mass in this regime. In the figure, we have denoted several benchmark mass values for dark matter and dark energy scenarios. We see that several dark matter scenarios, ranging from super-heavy WIMPzillas with $m_\varphi \sim 10^{13}$ GeV to WIMPs with $m_\varphi \sim$ GeV and to ultralight fuzzy dark matter with $m_\varphi \sim 10^{-22}$ eV, can be realized, as well as dark energy scenarios in which $m_\varphi \lesssim 10^{-33}$ eV. We will discuss these in further detail in the following sections. 

The extreme exponential sensitivity of the gravi-axion mass on the wormhole action is a generic result, but the above results for the GS wormhole are of course just one particular example. This wormhole solution assumes that the complex scalar is decomposed as in Eq.~\eqref{eq:phidecomp}. However, as discussed in Kallosh, Linde, Linde, and Susskind (KLLS) \cite{Kallosh:1995hi}, one can also consider a dynamical complex scalar field in which one decomposes $\Phi$ as 
\be 
\Phi = \bar{\mu}(r) e^{i\varphi/\mu},
\ee 
where $\mu$ remains the VEV of the field, but we have allowed field magnitude to vary as one approaches the wormhole solution and become comparable to $M_P$. In this case, one can approximate the wormhole action as \cite{Kallosh:1995hi} 
\be 
S_W^{\rm KLLS} \sim \ln(M_P/\mu),
\ee 
such that the full suppression factor of the mass is linear in $(\mu/M_P)$, rather than exponential. The wormhole throat radius in this case is \cite{Abbott:1989jw}
\be 
L_{\rm KLLS} \approx \frac{2}{\sqrt{3 \sqrt{2\pi}}}\frac{1}{M_P}.
\ee 
This Planckian size wormhole would ordinarily be subject to the  $\alpha^\prime$ corrections even in a weakly coupled string theory, but we follow \cite{Kallosh:1995hi} to explore this solution for demonstration purposes.  At first glance, this solution leads to unacceptably large masses for the gravi-axion; however, the authors of \cite{Kallosh:1995hi} also note that from string theory, one should expect an additional suppression by a factor of $\exp(-\pi^2/g_s)$, where $g_s$ is the string coupling. Taking this into account, we find 
\be 
m_{\varphi}^{\rm KLLS} \approx \frac{3\pi}{2\sqrt{2}}M_P\sqrt{\frac{M_P}{\mu}} e^{-\frac{8\pi^2}{g_s}}.
\ee 

\begin{figure}
    \includegraphics[width=0.5\textwidth]{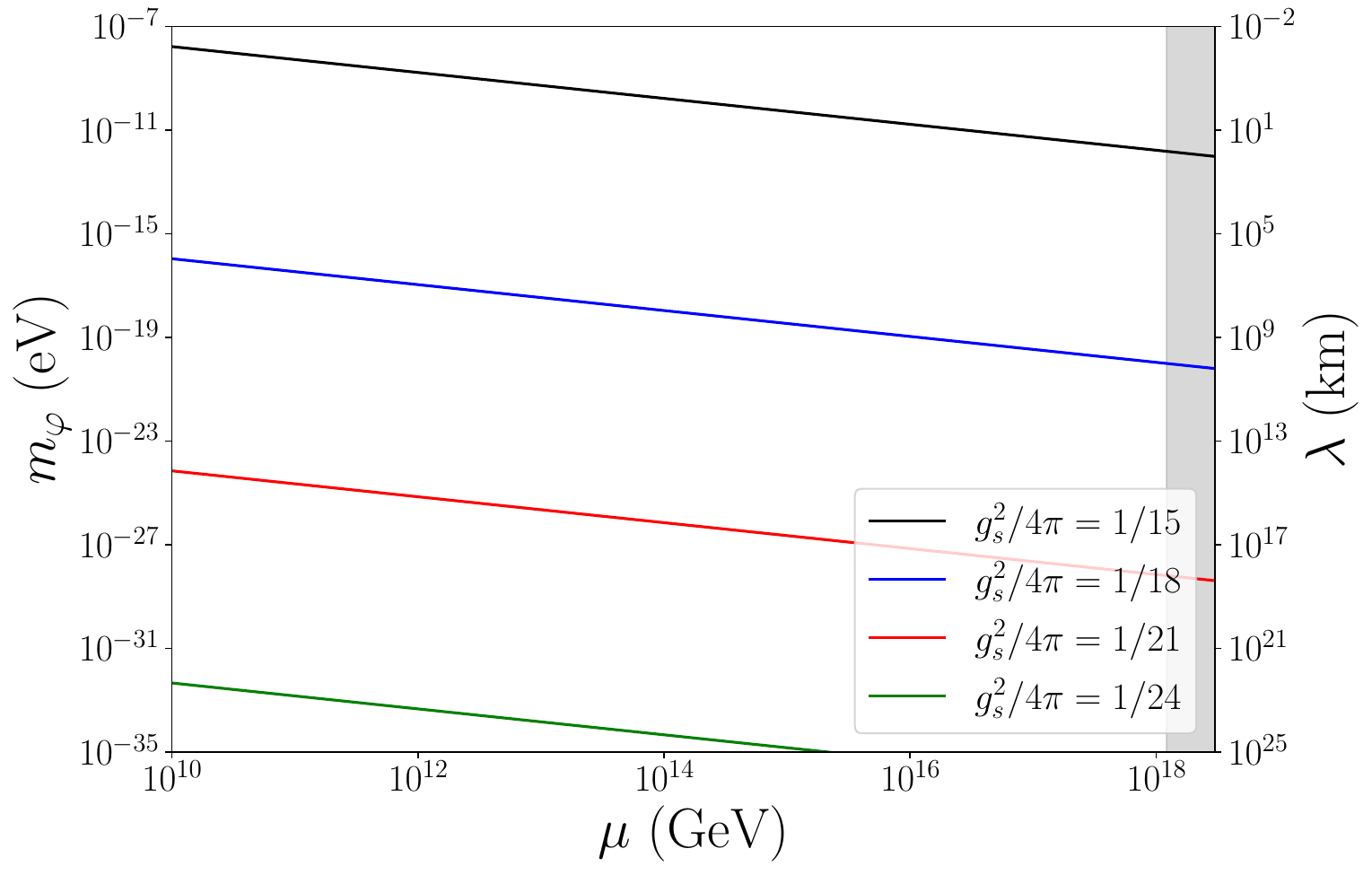}
    \caption{
    Mass of the gravi-axion in terms of the PQ scale, $\mu$, assuming the wormhole solution and stringy suppression discussed in \cite{Kallosh:1995hi}. The vertical gray shaded region denotes where the EFT condition, $\mu \ll M_p$ is no longer valid. We denote the compton wavelength, $\lambda$, of the gravi-axion on the right y-axis.
    }
    \label{fig:KLLS}
\end{figure}

In Figure~\ref{fig:KLLS}, we plot this realization for the gravi-axion mass for several choices of $g_s$, which depends on the string scale, but generically needs to satisfy $g_s^2/4\pi \lesssim 1/15$ \cite{Kallosh:1995hi}. 
We see that the dependence on $\mu$ is significantly less drastic than in the GS case, while the more important free parameter to set the gravi-axion mass is now the string coupling $g_s$. In this case, we can once again see that it is indeed possible to obtain gravi-axions in the ultralight dark matter or dark energy regimes, though in this scenario it is much more difficult to produce super-heavy dark matter, as compared to the $m_\varphi^{\rm GS}$ case. 

We can see that the GS and KLLS wormhole solutions admit gravi-axions in different, complementary, regimes. These two solutions arise from distinct physics; the GS wormhole from a general field theory, and the KLLS solution arising from string theory motivations. As discussed previously, the GS wormhole solution leads to a huge range in possible gravi-axion masses, but with a highly sensitive exponential dependence on $\mu$ in a narrow regime of parameter space. On the other hand, the KLLS solution, characterized by the lack of an exponential suppression due to $\mu$, leads to gravi-axions which are much less sensitive to the PQ scale, but are now exponentially sensitive to $g_s$. This leads to much lighter gravi-axions than what are possible in the GS scenario, but for a much wide range of $\mu$. It is of course possible that both of these classes of wormholes exist in nature and contribute to the gravi-axion mass in a non-trivial way. However, for simplicity, we focus on these two wormhole scenarios individually.

Before we move on, let us briefly comment on the size of the PQ scale as it pertains to the gravi-axion dCS coupling. From Eq.~\eqref{eq:LCS}, the dCS term is suppressed by a factor of $1/\mu$. In the early universe, we expect that the field values will be large such that this term gives a non-negligible contribution to the dynamics of the theory, while still remaining within the bounds of the EFT. However, we also are interested in the massive gravi-axion of the late universe, which perhaps could be probed with binary black holes mergers. In this case, the dCS term is generally parameterized as $\alpha\varphi R\tilde{R}/4$, where $\alpha = 1/(\mu M_P)$. For the values of $\mu$ necessary to get a cosmologically interesting gravi-axion mass, the coupling, $\alpha$, is highly suppressed. For example, if we take $\mu \sim 10^{16}$ GeV (in natural units), we have $\sqrt{\alpha} \sim 10^{-36}$ km (in geometric units $c = G = 1$). For comparison, the current best limit on $\alpha$ is $\sqrt{\alpha} \lesssim 8.5$ km \cite{Silva:2020acr}, which when saturated corresponds to $\mu\sim 10^{-59}$ GeV, and a gravi-axion mass of zero. This is of course well within the limit of the EFT, but is a vastly different scale than what is required for the mass generation. 
We then see clearly that if one wishes the gravi-axion to be cosmologically relevant, then the dCS coupling is exponentially supressed, while if one wishes the gravi-axion to lead to a dCS gravity theory at late times, then the mass of the dCS pseudoscalar is exponentially suppressed. That said, the super-low values for the PQ scale such as $\mu\sim 10^{-59}$ GeV, can't be motivated by any reasonable particle theory models. 

\section{Gravi-Axions as Dark Matter}
\label{sec:DM}

A natural application of gravi-axions is as dark matter. Pseudoscalar fields in a wide mass range are attractive candidates to account for some (or all) of the dark matter in the universe. Such fields could either be in the ultralight regime \cite{Hui:2021tkt}, in the canonical WIMP regime, or in the super-heavy regime as WIMPzillas \cite{Chung:1993ge}. As demonstrated in the last section, there is a reasonable parameter space to generate $\varphi$ with a mass to realize a viable dark matter scenario. We do not assume here that the gravi-axion has any additional couplings to SM particles, beyond its gravitational interactions, and therefore will not be thermally produced. In this section, we will consider two non-thermal production mechanisms and show how the correct dark matter relic abundance can be obtained. We first discuss production for ultralight gravi-axions via misalignment in Sec.~\ref{ssec:misalign}, and then discuss gravitational particle production of heavy gravi-axions in Sec.~\ref{ssec:GPP}. For simplicity, we will consider the mass induced by GS wormholes, hereafter $m_\varphi = m_\varphi^{\rm GS}$ as in Eq.~\eqref{eq:mGS}, but similar results will also hold for KLLS wormholes.

\subsection{Misalignment Production}
\label{ssec:misalign}

First, let us discuss production of gravi-axions via the misalignment mechanism. In this scenario, the initial value of the (rescaled) field, $\theta_i = \varphi_i/\mu$ is displaced from the minimum of its potential. As the universe expands, $m_\varphi$ becomes comparable to the decreasing Hubble parameter $H$, the field begins to roll and coherently oscillates about the minimum, leading to a condensate of stable axions that act as cold dark matter \cite{Kolb:1990vq}. As in Eq.~\eqref{eq:Seff-final}, we can approximate the potential once again as

\be 
V_\varphi \approx \frac{1}{2}m_\varphi^2 \varphi^2.
\ee 

The field is frozen at its initial value while $H \ll m_\varphi$, until 
\be 
H \approx m_\varphi,
\ee 
at which point the field will begin to oscillate. Then, the present-day abundance is given by~\cite{Marsh:2015xka, OHare:2024nmr}  
\be
\left(\frac{\Omega_\varphi h^2}{0.12}\right)_{\rm mis}\approx \theta_i^2\left(\frac{\mu}{4.7\times 10^{16} {\rm GeV}}\right)^2 \left(\frac{m_\varphi}{10^{-21} {\rm eV}}\right)^{1/2},
\label{eq:Relic-misalign}
\ee 
where $\Omega_\varphi h^2/0.12 \approx 1$ corresponds to the present-day dark matter abundance \cite{Planck:2018vyg}. To precisely determine the relic density, let us consider the scenarios in which the PQ symmetry breaking happens before and after inflation in turn.

\begin{figure}
    \includegraphics[width=\linewidth]{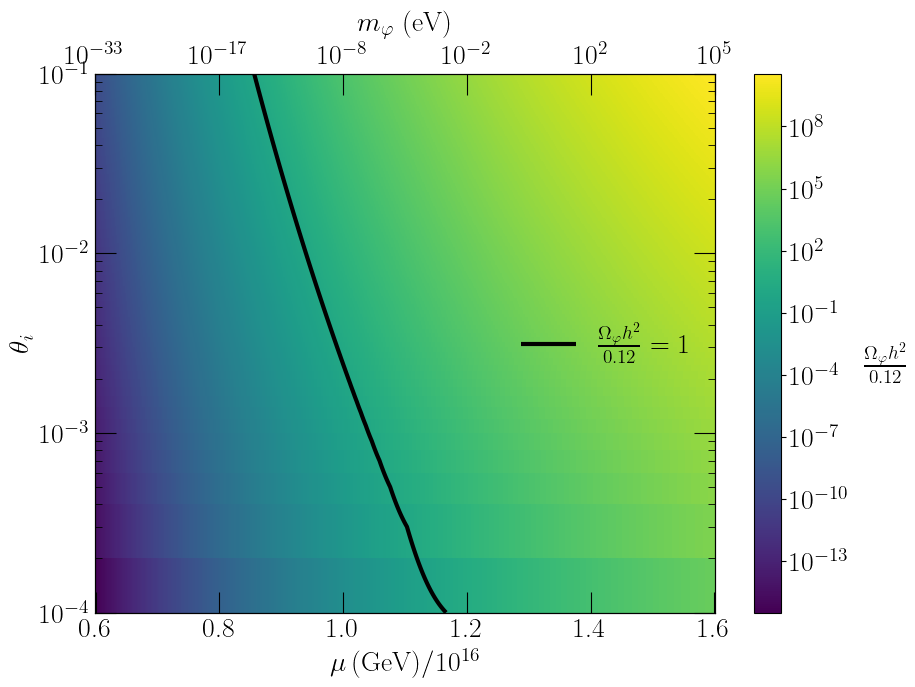}
    \caption{Gravi-axion abundance due to misalignment in a pre-inflationary symmetry breaking scenario, in terms of the initial misalignment angle, $\theta_i$ and symmetry breaking scale, $\mu$.}
    \label{fig:misalign-pre}
\end{figure}

\textbf{1. Pre-Inflationary Scenario:} First, let us consider the situation in which the symmetry breaking occurs before the end of inflation. In this case, the misalignment angle will be given by a single value, but is undetermined, thanks to the washout effects of inflation. From Eq.~\eqref{eq:Relic-misalign}, we can see that, within our regime of interest, it is feasible to produce the gravi-axion such that $\Omega_\varphi h^2/0.12 \approx 1$. Figure~\ref{fig:misalign-pre} shows the abundance in terms of $\theta_i$ and $\mu$. Very close to $\mu \sim 10^{16}$ GeV, we find the observed relic abundance~\cite{Planck:2018vyg}. 
For lower values of $\mu$, the gravi-axion is substantially under-produced and cannot give a non-neglible contribution to the dark matter abundance, while for $\mu > 1.2\times 10^{16}$ GeV, the initial misalignment angle has to be fine-tuned to be small, or the gravi-axion is overproduced. 

Notice that for the usual QCD axion or ALP scenario, the initial misalignment angle can take any value $-\pi < \theta_i < \pi$. However, in this case, because our theory is an EFT, we require that $\varphi/\mu \ll 1$ such that our calculation remains valid. This also allows us to neglect anharmonic effects from further terms beyond $m_\varphi^2\varphi^2$ when expanding the cosine potential. Finally, note that this scenario is also constrained by CMB measurements of near-vanishing isocurvature oscillations. However, these constraints can be evaded as long as \cite{OHare:2024nmr}
\be
\frac{H_e}{{\rm GeV}} \lesssim 10^7\theta_i \left(\frac{\mu}{10^{12} {\rm GeV}}\right),
\ee 
where $H_e$ is the Hubble parameter (which defines the energy scale) at the end of inflation.
Therefore, for initial misalignment angles of $\mathcal{O}(0.1)$ and $\mu \sim 10^{16}$ Gev, as long as $H_e \lesssim 10^{10}$ GeV, isocurvature is not an issue. 

\textbf{2: Post-Inflationary Scenario:} Let us now consider the possibility that the PQ symmetry breaking occurred after inflation. In this case, much of the above reasoning still holds, with some subtleties. First, the misalignment angle is no longer a free parameter, and is now characterized by its average value:
\be 
\langle \theta_i^2\rangle = \frac{\pi^2}{3}\,.
\ee 

Second, we now also have to take into account contributions from topological defects including cosmic strings and domain walls, which will lead to an enhanced abundance of gravi-axions beyond the density produced via misalignment alone \cite{Vilenkin:1984ib}. 

To characterize this more precisely, let us first consider cosmic strings. Cosmic strings are topological defects that can arise in cosmological phase transitions, when the axion winds around $2\pi$. Considering a gravi-axion field configuration in which every point has a randomly chosen angle, we expect that a network of these defects will arise. The cosmic strings will decay into gravi-axions, which lead to an enhancement of the abundance beyond misalignment alone. Similarly, we expect that symmetry breaking will also lead to the formation of domain walls, another class of topological defect, which are the boundary between different potential vacua. If more than one stable domain wall with high tension exists in the universe, they pose a problem for cosmology due to their potential for over-closing the universe. Since in our case the domain walls interpolate between physically identical vacua (the so-called $2\pi$ domain walls) they decay via quantum mechanical tunneling even 
if they are stable classically. Therefore, we will assume that the domain walls also collapse and decay into gravi-axions \cite{Vilenkin:1984ib}.

The enhancement in relic density due to topological defects depends on the specifics of the model and generally needs to be simulated numerically to fully characterize. Nonetheless, we can parameterize the contributions as $\alpha_{\rm cs}$ and $\alpha_{\rm dw}$ for cosmic strings and domain walls, respectively, such that the gravi-axion abundance are given by 
\be 
\left(\frac{\Omega_\varphi h^2 }{0.12}\right)_{\rm mis, \,post} \approx \left(\frac{\Omega_\varphi h^2 }{0.12}\right)_{\rm mis}(1 + \alpha_{\rm cs} + \alpha_{\rm dw}),
\ee 
where $\left({\Omega_\varphi h^2 }/{0.12}\right)_{\rm mis} $ is the abundance from Eq.~\eqref{eq:Relic-misalign}, setting $\langle \theta_i^2\rangle = \pi^2/3$. Depending on the magnitude of these contributions, it is possible to obtain an enhancement in the gravi-axion abundance, as compared to the pre-inflationary misalignment scenario shown in Fig.~\ref{fig:misalign-pre}.

\subsection{Gravitational Particle Production}
\label{ssec:GPP}

If the symmetry breaking occurs before inflation, the gravi-axion particle abundance can also be produced via cosmological gravitational particle production (GPP) \cite{Parker:1969au, Parker:1971pt, Ford:2021syk,Chung:1998zb,Chung:1998ua, Kolb:2023ydq}. In this process, particles are produced due to adiabaticity violation during the rapid expansion of spacetime in inflation. This is an old idea, going back to Schr\"odinger \cite{SCHRODINGER1939899}, and has since been studied in great detail for a variety of different inflationary scenarios and fields \cite{Graham:2015rva, Ema:2015dka, Ema:2016hlw, Ema:2019yrd, Kolb:2020fwh,Ahmed:2020fhc, Alexander:2020gmv,Kolb:2021xfn,Kolb:2021nob,Kolb:2023dzp, Capanelli:2023uwv,Kaneta:2023uwi,Maleknejad:2022gyf,Capanelli:2024rlk,Capanelli:2024pzd,Jenks:2024fiu,Racco:2024aac, Chowdhury:2025mye}. In general, this mechanism can efficiently produce particles with masses $m_\varphi \sim \mathcal{O}(H_e)$, recalling that $H_e$ is the Hubble parameter at the end of inflation. However, depending on the specific scenario, this mechanism can also produce lighter particles.  Figure~\ref{fig:n1mass} shows that, for a wide range of inflationary models, gravitational particle creating can be a relevant production mechanism. For example, if we take a generic high-scale for the end of inflation at $H_e\sim 10^{13}$ GeV, a symmetry breaking scale of $\mu \sim 5 \times 10^{16}$ GeV can easily produce particles with masses of $\mathcal{O}(H_e)$.
Our model, with symmetry breaking at high energies, also provides an elegant explanation for why an additional spectator field would be present during the inflationary epoch, which is a necessary component of the GPP mechanism.

We can easily determine the particle abundance generated from this mechanism. In our case, $\varphi$ is non-minimally coupled to gravity via the $(\varphi/\mu) \tilde{R}R$ term. However, on an Friedmann-Lemaitre-Robertson-Walker (FLRW) background $\tilde{R}R$ vanishes (irrespective of the value of $\varphi/\mu$), and we can therefore treat the gravi-axion as a minimally coupled scalar field, which has been well-studied in the context of GPP (see the review \cite{Kolb:2023ydq} and references therein). Consider then a cosmological FLRW spacetime, defined by $g_{\mu\nu} = a^2(\eta){\rm diag}(1,-1,-1,-1)$ in conformal coordinates (and with a mostly-minus metric signature), where $\eta$ is conformal time. In order to canonically normalize the kinetic term, we define $\bar{\chi} (\eta, \mathbf{x}) = a(\eta)\varphi(\eta, \mathbf{x})$. The field $\bar\chi$ can then be decomposed into the Fourier modes: 
\begin{equation}
   \bar{\chi}(\eta,\textbf{x})=\int\frac{d^3\textbf{k}}{(2\pi)^3}\left[\hat{a}_{\textbf{k}}\chi_{\textbf{k}}(\eta)e^{i\textbf{k}\cdot \textbf{x}}+\hat{a}_{\textbf{k}}^{\dagger}\chi_{\textbf{k}}^*(\eta)e^{-i\textbf{k}\cdot \textbf{x}}\right],
\end{equation}

where $k = |\textbf{k}|$ is the comoving wavenumber and $\hat{a}_k^{\dagger}$ and $\hat{a}_k$ are the creation and annihilation operators respectively. The rescaled gravi-axion satisfies the following mode equation:
\begin{equation}
    \partial_{\eta}^2\chi_k(\eta)+\omega_k^2(\eta)\chi_k(\eta)=0,
    \label{eq:mode_eq}
\end{equation}
where 
\be 
\omega_k^2 = k^2 + a^2(\eta) m_\varphi^2.
\ee 

From the mode functions, one can determine the spectrum of particles produced:
\begin{equation}
\label{eq:nk}
    a^3 n_k= \lim_{\eta \rightarrow \infty} \frac{k^3}{2\pi^2}|\beta_k|^2,
\end{equation}
which is constructed in terms of the Bogoliobov coefficient, $\beta_k$, where
\begin{equation}
    |\beta_k|^2=\frac{1}{\omega_k}\left(\frac{1}{2}|\partial_{\eta}\chi_k|^2+\frac{1}{2}\omega_k^2|\chi_k|^2\right)-\frac{1}{2}~.
\end{equation}
Then, the comoving particle number density at the end of inflation is given by
\begin{equation}
     a^3 n = \int a^3 n_k \, d\log k. 
     \label{eq:na3}
\end{equation}

From the comoving number density, the late-time relic density can be determined, depending on when reheating occurred. For simplicity, let us consider an instantaneous reheating scenario. In this case, the number density is given by \cite{Kolb:2023ydq}
\be 
\frac{a^3 n}{a_e^2 H_e^3} = \frac{1}{8\pi^2}\left(\frac{H_e}{m_\varphi}\right)^{1/2}\log\left(\frac{k_\star}{k_{\rm min}}\right),
\ee 
where $a_e$ is the scale factor at the end of inflation, $k_\star$ corresponds to the conformal wavenumber of the modes that leave the horizon during inflation and reenter when the Hubble parameter is order the field mass, and $k_{\rm min}$ is the conformal wavenumber associated with an IR cutoff imposed to regulate the $n a^3$ integral in Eq.~\eqref{eq:na3}. In this scenario, we can take $k_\star/{(a_eH_e)} \approx (m_\varphi/H_e)^{1/2}$ and $k_{\rm min}\approx a_0H_0$, where $a_0$ and $H_0$ are the present day values of the scale factor and the Hubble parameter, respectively. From here, we can approximate the current energy density in the field as 
\begin{align} 
\frac{\Omega_\varphi h^2}{0.12} \approx &\left(\frac{H_e}{10^{14} {\rm GeV}}\right)^2 \left(\frac{m_\varphi}{10^{-5} {\rm eV}}\right)^{1/2}\nonumber \\
&\times\left(\frac{g_{\star, \rm{R}}}{106.75}\right)^{-1/4} \log\left(\frac{k_\star}{k_{\rm min}}\right),
\end{align} 
where $g_{\star, {\rm R}}$ characterizes the effective number of degrees of freedom in the plasma at reheating. Taking $g_{\star, {\rm R}} = 106.75$, we are left with $H_e$ and $m_\varphi$ as free parameters. 

\begin{figure}
    \includegraphics[width=\linewidth]{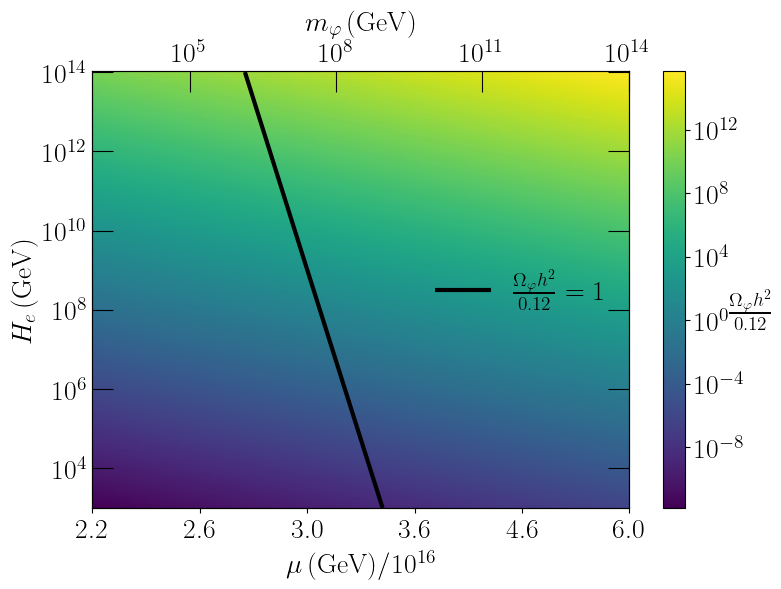}
    \caption{Abundance of gravi-axions produced via gravitational particle production for a range of masses and inflationary scales.}
    \label{fig:GPP}
\end{figure}

Figure~\ref{fig:GPP} shows the energy density in terms of $m_\varphi$ and $H_e$. We remain agnostic to the scale of inflation such that it could range from low to high scales, and therefore consider masses from $10^3 < m_\varphi/{\rm GeV} < 10^{14}$. The black line shows the contour such that $\Omega_\varphi h^2 = 0.12$ to match the observed present-day abundance~\cite{Planck:2018vyg}. Clearly for much of this parameter space, the gravi-axion is dramatically overproduced, particularly with high scale inflation. However, there is still a wide range of parameter space where this scenario is viable. Furthermore, note that our instantaneous reheating approximation provides an upper limit on the particle production, and considering later reheating scenarios will also shift the $\Omega_\varphi h^2=0.12$ contour to higher $H_e$. 

Another way to appreciably widen the acceptable parameter space is to consider the decay of the gravi-axion into another particle, which could then be the dark matter: $\varphi \rightarrow \varphi'\varphi'$, where now $\varphi'$ is the dark matter. In this case, the abundance will be given by 
\be 
\frac{\Omega_{\varphi'} h^2}{0.12}= 2 \left(\frac{m_\varphi'}{m_\varphi}\right) \left(\frac{\Omega_{\varphi}h^2}{0.12}\right),
\ee 
which clearly allows for a large range of values of $m_\varphi$ and $H_e$, provided that the decay mass is sufficiently light. 

Finally, we note that the gravitational particle production of minimally coupled scalars is constrained by isocurvature oscillations, as with pre-inflationary misalignment. In general, as long as $m_\varphi \gtrsim H_e$, then these constraints can be evaded and the above dark matter scenarios can be successfully realized. 

\section{Gravi-Axion Decay}
\label{sec:decay}

Given that the gravi-axion couples directly to gravity and assumes no additional couplings to Standard Model particles, it is useful to consider unique observational windows that are distinct from other axion-like dark matter candidates. A natural application is to gravitational waves generated in the early universe. In particular, due to the direct coupling between the gravi-axion and gravity, an intriguing possibility is that some fraction of the produced gravi-axions can decay to gravitons and leave behind a distinctive gravitational-wave background (GWB) signature if $\varphi/\mu$ is large enough. As discussed in Sec.~\ref{sec:DM}, for some of the viable parameter space, the gravi-axion abundance is overproduced to account for the present-day cosmological dark matter fraction. If a large fraction of the gravi-axion could decay into gravitons, it could be possible to leave a characteristic GWB signature, while maintaining the correct dark matter abundance. A similar scenario has been considered recently for dark matter axion decay in our galaxy \cite{Figliolia:2025dtw}, but we will now consider this decay occuring in the early universe, as was briefly mentioned in \cite{Ema:2021fdz}. This procedure is distinct from other models of axions and ALPs from U(1) symmetry breaking, because our model has a direct coupling of the gravi-axion to gravity, such that it can decay directly to gravitons even when other axions cannot. 

Let us then consider the decay of the axion into two gravitons, $\varphi \rightarrow hh$ as is allowed by the coupling between $\varphi$ and $R\tilde{R}$. In this case, the decay rate is \cite{Ema:2021fdz}
\be 
\Gamma_{\varphi\rightarrow hh} = \frac{1}{4\pi}\frac{m_\varphi^7}{\mu^2 M_P^4}.
\ee 
For light gravi-axions, the decay rate is clearly highly suppressed, and therefore the lifetime, $\tau \sim \Gamma^{-1}$, is extremely large, making an appreciable present-day gravitational-wave signal unlikely. However, for heavier gravi-axions, such as those produced via GPP, $\Gamma$ can be large enough to produce a non-negligible background. This is the scenario that we will focus on for the rest of this section. 

The gravitational-wave energy spectrum in terms of the GW energy density, $\rho_h$ and the GW energy, $E$, is given by \cite{Ema:2021fdz}

\be 
\frac{d\rho_h}{d\ln E} = E^2 \int \frac{dz}{H(z)}\Gamma_{\varphi \rightarrow hh}n_\varphi(z) a^3(z) \frac{dN_h}{dE'},
\ee 
where $n_\varphi$ is the number density of gravi-axions before they begin to decay, $H(z)$ is the Hubble parameter, and the number of gravitons produced per decay per unit energy, $dN_h/dE' $, is defined as
\be 
\frac{dN_h}{dE'} = 2\delta \left[E' - \frac{m_\varphi}{2}\right],
\label{eq:dNdEp}
\ee 
where $E' = (1+z)E$ is the energy at which the gravi-axion decays.
The integral can be done analytically to obtain 
\be 
\frac{d\rho_h}{d\ln E} = \frac{16 E^4}{m_\varphi^2}\frac{\Gamma_{\varphi\rightarrow hh}}{H(z_d)}\left(\frac{\rho_\varphi}{s}\right) s(z_d)e^{-\Gamma t(z_d)},
\ee 
where $\rho_\varphi/s$ is the energy density in the gravi-axions before they begin to decay, $s(z)$ is the entropy density at a temperature $T$, given by 
\be 
s(z) = \frac{2\pi^2}{45}g_{\star,s}(T)T^3,
\ee 
and we have assumed that $\Gamma_{\varphi\rightarrow hh}$ is the only decay channel. The redshift at which the gravi-axions decay, $z_d$, is defined from Eq.~\eqref{eq:dNdEp} by 
\be 
z_d = \frac{m_\varphi}{2E} - 1.
\ee 

\begin{figure}
    \includegraphics[width=0.45\textwidth]{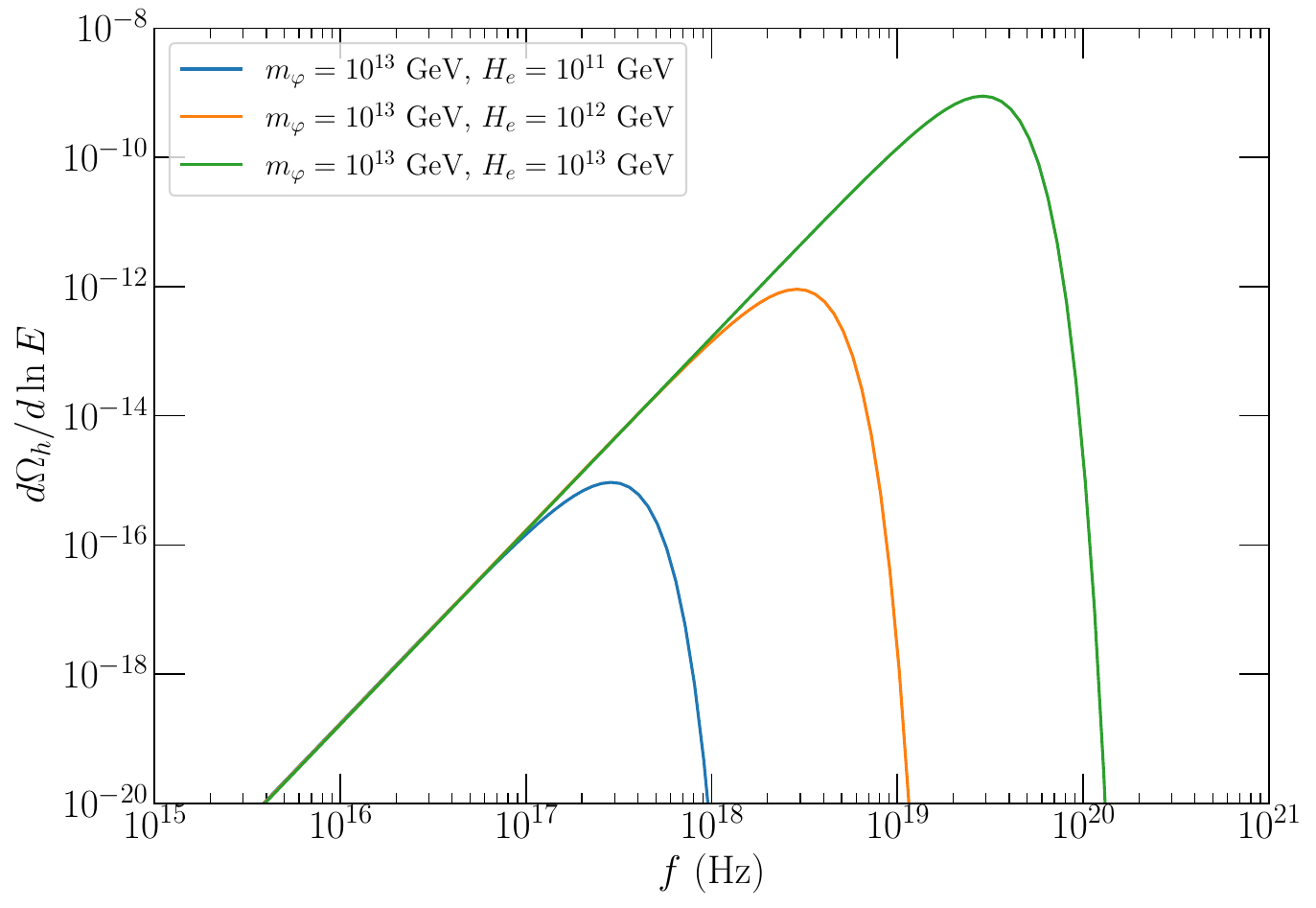}
    \caption{Gravitational-wave spectrum from gravi-axion decay for three fiducial examples of a gravi-axion population produced via GPP.}
    \label{fig:GWs}
\end{figure}

With this in hand, and assuming instantaneous reheating as in Sec.~\ref{ssec:GPP}, and that the gravi-axions decay during radiation domination, we can determine the spectrum of gravitational waves. Let us consider then
\be
\frac{d\Omega_h}{d\ln E} = \frac{1}{\rho_{\rm crit}}\frac{d\rho_h}{d\ln E},
\ee 
where $\rho_{\rm crit}$ is the critical density of the universe today. Figure~\eqref{fig:GWs} shows the expected gravitational-wave spectrum from gravi-axion decay for three examples of the GPP production parameter space that give rise to overproduction. We consider a gravi-axion with $m_\varphi = 10^{13}$ GeV, and the GWB produced for $H_e=10^{11}, 10^{12}, 10^{13}$ GeV. Observe that, indeed, a large amplitude of gravitational waves can be created from gravi-axion decay, albeit at frequencies higher than what can currently be probed with existing ground-based detectors, such as advanced LIGO, Virgo and KAGRA. The amplitude of the GWB increases with the inflationary Hubble scale, as does the peak frequency. While this frequency range is unattainable with current detectors, there are several proposals to probe high-frequency ($f \sim {\rm GHz}$) gravitational waves with resonant cavity experiments, e.g. \cite{Berlin:2021txa}. 

In the above, we have assumed that the full abundance of gravi-axions has decayed to gravitons. However, an interesting possibility is that, in the regime of parameter space where the gravi-axion is overproduced, some large fraction could decay such that a gravi-axion population remains to be the dark matter. Then, these gravitational waves would be a smoking gun signature of gravi-axion dark matter.

\section{Gravi-Axions as Dark Energy}
\label{sec:DE}

As discussed Sec.~\ref{sec:dcswormholes} and shown in Fig.~\ref{fig:n1mass}, gravi-axions can also be produced in a mass range relevant to dark energy. If $\mu$ is on the lower end of the viable mass-generation region, then the gravi-axion will resemble a quintessence field. Quintessence models are characterized by a dynamical scalar field that is generally minimally coupled to gravity \cite{Frieman:1995pm, Chiba:2009sj,Dutta:2008qn,Linder:2007wa,Carroll:1998zi}. To reach agreement with the present-day cosmic acceleration, the mass of the quintessence scalar must be smaller than the Hubble rate today, $m_\varphi \lesssim H_0 \sim 10^{-33} {\rm eV}$ \cite{Tsujikawa:2013fta}. In our scenario, the gravitational axion mass can naturally be generated in the required ultralight regime, due to the exponential suppression by $S_W$ of $m_\varphi$. This is the gravitational analog to generating a quintessence axion potential via electroweak instantons, as discussed in \cite{Nomura:2000yk}. 

Traditionally, quintessence models are defined by only a scalar field minimally coupled to gravity,.\footnote{Note that \cite{Smith:2007jm} suggested that a quintessence field could couple to the dCS term. Similarly, quintessence has been studied with a gravitational coupling to the Gauss-Bonnet curvature invariant (the parity-even analog of the dCS term) \cite{Koivisto:2006ai}, which has since been ruled out from the binary neutron star observation GW170817  \cite{Ezquiaga:2017ekz}.} but we can easily see that the addition of the dCS coupling does not actually change the dynamics in this case. As mentioned before, on an FLRW background, $\tilde{R}R$ vanishes due to the spherical symmetry of the spacetime, leaving the field equations unchanged from the minimally coupled scenario: 
\be 
\ddot{\varphi} + 3 H \dot{\varphi} + \frac{dV_\varphi}{d\varphi} =0,
\ee 

and the equation of state is 
\be 
w = \frac{\dot{\varphi}/2 -V_\varphi(\varphi)}{\dot{\varphi}/2 +V_\varphi(\varphi)}.
\ee
Therefore, on cosmological scales, the coupling to $R\tilde{R}$ can be neglected, leaving $\varphi$ to behave like a standard quintessence field.

There are a variety of potentials that have been explored to realize the quintessence scenario. Our model typically generates a cosine potential of the form $V_\varphi \propto \cos(\varphi/\mu)$, as in Eq.~\eqref{eq:Seff1}, similar to a pseudo-Nambu-Goldstone Boson quintessence model \cite{Frieman:1995pm}. Our potential can also be expanded as in the previous section  to obtain a potential of the form $V_\varphi \propto m^2\varphi^2$ as well as a $\lambda \varphi^4$ term. Models with these potentials can exhibit thawing behavior, in which the field freezes at early times due to Hubble friction, acting like a cosmological constant. The field begins to evolve when $m_\varphi \sim H$, `thawing' away from $w=-1$. 

Quintessence models with thawing behavior are of particular interest due to the recent results from the Dark Energy Spectroscopic Instrument (DESI). Cosmological observations with the latter have indicated a $\sim 3\sigma$ preference for evolving dark energy at late times, away from the standard $\Lambda$CDM paradigm of a cosmological constant with $w_\Lambda = -1$ \cite{DESI:2025zgx}. Thawing quintessence models have been shown to be a better fit to the DESI data than $\Lambda$CDM~\cite{DESI:2025fii, Shajib:2025tpd}. Therefore, the gravitational axion model is a timely candidate for quintessence that could be relevant for these new cosmological data.

Before we move on, let us briefly comment on the validity of our EFT as it pertains to the necessary field values to realize the gravi-axion as a viable quintessence model. The present day energy density of dark energy is measured to be $\rho_{\rm DE} \approx 10^{-12} {\rm {eV}}^4$. For a gravi-axion of mass $m_\varphi \sim 10^{-33} {\rm eV}$, the required field value to obtain this energy density would be $\varphi \sim 10^{27}$ eV, which is near $M_P$ and leads to $\varphi \gg \mu$, such that the $m_\varphi^2\varphi^2$ term dominates over the others. As this term is generated as a low-energy limit of the original high-energy action, one may worry about contributions from higher-order terms in the potential. However, as can be appreciated from Eq.~\eqref{eq:Seff1}, higher-order contributions will also come with additional suppressions by the wormhole factor, $g_n \sim e^{-S_W}$, where $S_W$ has some dependence on $n$. Due to the exponential dependence on $S_W$, the next-order terms will be sufficiently suppressed to neglect their inclusion.

\section{Discussion and Conclusions}
\label{sec:conclusions}

In this paper we have discussed the cosmological implications of a theory of gravi-axions. In our setup, the gravi-axion arises as an effective theory from a a weakly coupled UV theory of a massive, complex scalar field coupled to chiral fermions, with an unbroken U(1) symmetry. First, we consider the generation of a potential for the gravi-axion due to non-perturbative wormhole effects, explicitly breaking the symmetry to its discrete subgroup of $2\pi n$ shifts. Then, the gravi-axion coupling to $R\tilde{R}$ arises from integrating out the massive fermions modes. We are left with a theory of a massive gravi-axion, which breaks the continuous shift symmetry of dCS gravity. 

Our current model of gravi-axions mass generation is highly dependent on the wormhole solution one considers to capture non-perturbative effects. Though the true wormhole solution that Nature gives us depends on unknown high-energy physics, we have shown here that, as a proof of principle, it is possible to obtain the gravi-axion mass through non-perturbative effects in a cosmologically relevant range, depending on the scale of $\mu$ for both the GS and KLLS wormholes. We then discussed the phenomenology of the gravi-axion as a dark matter candidate, showing how it can be produced for a wide range of masses via misalignment or gravitational particle production, in order to account for the relic density of dark matter today, depending on the PQ scale of the theory. We calculated the decay of gravi-axions into gravitons, suggesting that a contribution to the gravitational-wave background could be a hallmark signature of this model. Finally, for lower values of the PQ scale, we showed that the gravi-axion mass can be in the right range to act as dynamical dark energy, necessitating further study.  

As always, there are many caveats that generate directions for further study. A particularly intriguing possibility is that the same mechanism responsible for the generation of our gravi-axion theory is also related to the leptogenesis mechanism suggested in \cite{Alexander:lepto}. In that scenario, the necessary scale for successful leptogenesis is near $\mu \sim M_P$. While $\mu \approx M_P$ is not within the regime of validity of our effective theory, it is possible that the very high scale of symmetry breaking at $\mu \sim 10^{16}$--$10^{17}$ GeV may play an important role. Similarly, Ref.~\cite{Alexander:2014bsa} has suggested that preheating could be realized for a scenario in which the inflaton itself has a dCS coupling. Further study could investigate if the gravi-axion could provide an analogous role. Aside from leptogenesis, it is well known that the gravi-axion term by itself can lead to birefringence of gravitational waves if $\dot{\varphi}$ is non-negligible. It would be interesting to explore how gravitational waves generated in the early universe may inherit some chirality as a result, and how this may impact other observables, such as the cosmic microwave background \cite{Inomata:2018rin, Inomata:2018vbu} or large scale structure \cite{Inomata:2024ald}. Finally, it would be useful to determine if beyond our semi-classical approximation, there are additional contributions to generate a massive gravi-axion with a non-negligible coupling to the Pontryagin density at late times. We leave these and other interesting questions for future work.

\acknowledgements
SA acknowledges support from the Simons Foundation, Award 896696. GG was supported in part by the NSF grant PHY-2210349. LJ is supported by the Provost's Postdoctoral Fellowship at Johns Hopkins University and was supported by the Kavli Institute for Cosmological Physics at the University of Chicago while a portion of this work was carried out. NY acknowledges support from the Simons
Foundation through Award No.~896696, the Simons Foundation International through Award No.~SFI-MPS-BH-00012593-01, the NSF through Grants No.~PHY-2207650 and~PHY-25-12423, and NASA through Grant No. 80NSSC22K0806.

\appendix

\section{Review of Euclidean Wormholes}
\label{ssec:instantons}

Axions and axion-like particles are known to obtain a mass from non-perturbative symmetry breaking by gravitational instantons or Euclidean wormhole solutions at high energies. In this Appendix, we briefly summarize the basic mechanics of this procedure within the usual framework. We would like to determine the contribution to the low-energy effective action arising from such non-perturbative wormhole effects. In the following, we will closely follow the discussion in \cite{Hebecker:2018ofv}, but we also point the interested reader to the original works \cite{Lee:1988ge, Rey:1989mg, Abbott:1989jw}. 

Euclidean wormholes are non-perturbative fluctuations of the topology of spacetime that arise at high energy scales. These wormholes can carry axion charge, and therefore are important to take into account when computing the effective axion theory at low energies. In particular, they need to be accounted for when computing the path integral. Schematically, each wormhole is characterized by the action $S_W$, which then contributes a factor of $e^{-S_{W}}$ to the path integral. This leads to an overall breaking of the continuous axion shift symmetry to a discrete subgroup and generates a potential for the axion, which is proportional to $e^{-S_W}$. Physically, we can think of this breaking of the continuous shift symmetry as a result of the wormholes `eating' or sourcing the axion charge. Below, we sketch out the technical calculation which leads to the well-known axion cosine potential.

The partition function describing wormhole topologies can be written as
\begin{align} 
Z_W &= \int Dg D\phi \; e^{-S[g,\phi]}\nonumber \\
&\times\left(\int d^4x \sqrt{g(x)}\int d^4y \sqrt{g(y)}e^{-S_W[x,y,g,\phi]}\right),
\label{eq:Z1}
\end{align} 
where $S[g, \phi]$ is the background action, containing standard Einstein-Hilbert and matter terms, and $S_W[x,y,g,\phi]$ is the correction to the action from inserting a wormhole that connects the spacetime points $x, y$ in the background, and where $Dg$ and $D\phi$ are the integral measures over all spacetime metrics on the background universe and over all additional fields, respectively. Restricting attention to soft (low-energy compared to $M_P$) field and metric configurations, the wormhole action, $S_W$, can be expanded generically in the bilocal expression 
\be 
S_W = \bar{S}_W + \sum_{ij} \tilde{\Delta}_{ij}\mathcal{O}_i(x)\mathcal{O}_j(y),
\ee 

where $\bar{S}_W$ is the wormhole action on the unperturbed background and $\mathcal{O}_i$ are bilocal operators. The above can be exponentiated to give 
\be 
e^{-S_W} = \frac{1}{2}\Delta_{ij}\mathcal{O}_i(x)\mathcal{O}_j(y),
\ee 
where we have absorbed the background wormhole action into the exponential factor $\Delta_{ij} \propto e^{-\bar{S}_W}$. 

Using this expression and the dilute gas approximation, assuming that the inter-wormhole distance is much larger than the throat diameter, for $n$ wormholes, the partition function in Eq.~\eqref{eq:Z1} becomes 

\be 
Z_W = \int Dg D\phi \; e^{-S[g, \phi] + I}, 
\ee 

where the bilocal action, I, is
\be 
I = \frac{1}{2} \int d^4 x \sqrt{g} \int d^4 y \sqrt{g} \sum_{i,j} \Delta_{ij}\mathcal{O}_i(x)\mathcal{O}_j(y).
\ee 
To give the action $I$ a local form, one can introduce auxiliary parameters $\alpha_i$ to write the action 
\begin{align} 
e^I &= \Pi_i\left(\int \frac{d\alpha_i}{\sqrt{2\pi}}\right) {\rm exp}\left(-\frac{1}{2} \sum_{i,j} \alpha_i \Delta_{ij}^{-1}\alpha_j\right) \nonumber \\
&\times{\rm exp}\left(\sum_{i} \alpha_i \int d^4x \sqrt{g}\mathcal{O}_i(x)\right),
\end{align} 
where $1/\sqrt{2\pi}$ has been introduced as a normalization factor.
If one thinks about the contributions from wormholes as `baby universes' that are emitted and absorbed by some larger `parent universe,' one can write the above as 
\be 
e^I = \int \frac{d\alpha}{\sqrt{2\pi}} {\rm exp}\left(-\frac{1}{2}\alpha^2 + \alpha\sqrt{\Delta}\int d^4 x \sqrt{g}\mathcal{O}(x)\right),
\label{eq:wormholeeI}
\ee 

where we have simplified by considering only a single operator and have rescaled $\alpha \rightarrow \alpha\sqrt{\Delta}$. Analogous to the harmonic oscillator, we will define the state with no baby universes as the vacuum, $\ket{0}$, such that a state with $n$ baby universes is 
\be 
\ket{n} = \frac{(a^\dagger)^n}{\sqrt{n!}} \ket{0},
\ee 
where the parameter $\alpha$ is the eigenvalue of the operator  $\hat{\alpha} = a + a^\dagger$, so that $\hat{\alpha}\ket{\alpha} = \alpha \ket{\alpha}$. We can then write Eq.~\eqref{eq:wormholeeI} as
\be 
e^I = \bra{0}(a + a^\dagger)\mathcal{O}\ket{0}, 
\ee 
where
\be 
\bra{0}(a + a^\dagger)^n\ket{0} = \int \frac{d\alpha}{\sqrt{2\pi}} {\rm exp}
\left(-\frac{1}{2}\alpha^2\right) \alpha^n.
\ee 

This expression can easily be generalized to arbitrary initial and final states and number of operators to obtain 
\be 
e^I = \bra{\psi_2}{\rm exp}\left[\sum_i\sqrt{\Delta_{ii}}\int d^4x \sqrt{g}\mathcal{O}_i(x)(a_i + a_i^\dagger)\right]\ket{\psi_1}.
\ee 
The matrix $\Delta_{ii}$ can be decomposed into a contribution from the wormhole action, $e^{-S_W}$ and a remainder, which we will denote as $K_n$. 

The local operators have the form $\mathcal{O}_i = e^{in\theta/f_a} \mathcal{O}_S$, where $\mathcal{O}_S$ is a singlet operator, $\mathcal{O}_S =  1 + a \mathcal{R} + ...$. The action itself can then be written as 
\be 
I = \sum_n e^{-\frac{S_W}{2}}K_n \int d^4 x \sqrt{g(x)}\mathcal{O}_S \alpha_n \exp{\frac{ i n\phi}{f_a}},
\ee 
where the factor of $S_W/2$ arises because each wormhole can be thought of as an instanton/anti-instanton pair, each contributing $S_W/2$. Taking the unit operator of $\mathcal{O}_S$ and taking into account the Hermitian conjugate of $I$ will finally yield the correction to the action
\be 
I = \sum_n e^{-S_W/2}K_n \int d^4 x \sqrt{g(x)}\alpha_n \cos\frac{n\phi}{f_a},
\ee 
corresponding to a potential for $\phi$.

From this, we obtain the correction to the low-energy action from the wormholes. The constant $K_n$ is expected to be $K \propto 1/L^4$, where $L$ is the length of the wormhole throat,
but its actual value cannot generally be computed \cite{Alonso:2017avz}. The key feature of this expression is the exponential dependence on the wormhole action, $e^{-S_W}$, which will be the dominant contribution in determining the size of the correction. If one takes into account only wormholes with charge $n=\pm 1$, which will dominate the contribution when $S_W$ is large, the potential has the usual axion form.
\be 
V_\varphi \sim e^{-S_W/2} \frac{\alpha_1}{L^4}\cos\frac{\phi}{f_a}.
\ee 

 In the above, we have left the discussion generic and not specified to any particular wormhole solution. As a particular example though, it has been shown that a theory of complex scalars will admit wormhole solutions. This was first shown in the dual formulation in \cite{Giddings:1987cg}, and then when the scalar theory is written explicitly \cite{Abbott:1989jw, Lee:1988ge}, which is the scenario we have considered in this work.

\section{The Giddings-Strominger Wormhole}
\label{ssec:GS}
In this appendix, we review the Giddings-Strominger wormhole solution. We follow \cite{Giddings:1987cg} as well as the discussions in \cite{Alonso:2017avz, Hebecker:2018ofv}. Let us begin with the Euclidean action for a three-form field strength, $H = dB$ minimally coupled to gravity:
\be 
S = \int d^4x \sqrt{g}\left(-\frac{M_{P}^2}{16\pi}R + \frac{\mathcal{F}}{2}H_{\mu\nu\kappa}H^{\mu\nu\kappa}\right), 
\label{eq:GSaction-app}
\ee 
where $M_P$ is the Planck mass, and $\mathcal{F}$ is related to the Peccei-Quinn scale, $f_a$ by $\mathcal{F} = {1}/({3! f_a^2)}$. The three-form H is related to the axion via the duality 
\be 
H = f_a^2 \epsilon_{\mu\nu\rho\varphi}\partial^\rho\theta,
\ee 
which allows one to equivalently write the action in terms of the axion. 

Varying the action, Eq.~\eqref{eq:GSaction-app} with respect to the metric, one can extract the Einstein equations:
\be 
R_{\mu\nu} - \frac{1}{2}g_{\mu\nu}R = \frac{8\pi}{M_{P}^2}\mathcal{F}T_{\mu\nu}, 
\label{eq:GSEE}
\ee 
where the stress-energy tensor, $T_{\mu\nu}$ is
\be 
T_{\mu\nu} = 3H_{\mu\alpha\beta}H_\nu^{\alpha\beta} - \frac{1}{2}g_{\mu\nu}H^2. 
\ee 
It is useful to write Eq.~\eqref{eq:GSEE} in trace form to obtain 
\be 
-R = \frac{8\pi}{M_P^2}\mathcal{F} H^2. 
\ee 
Varying the action with respect to $B$ yields the equation of motion:
\be 
d\star H = 0. 
\ee 

Now consider a spherically-symmetric spacetime ansatz, such that 
\be 
ds^2 = \alpha^2(\tau)dr^2 + \beta^2(\tau)d\Omega_3^2, 
\ee 
and
\be 
H_{\mu\nu\rho} = \mathcal{H}(\tau)\epsilon_{ijk} \delta^i_\mu \delta^j_\nu \delta^k_\rho,
\ee 
where $\tau$ is now the Euclidean time coordinate. The $(\tau,\tau)$ component of the Einstein equations is then
\be 
-\frac{3k\alpha^2}{\beta^2} + \frac{3\dot{\beta}^2}{\beta^2} = \frac{8\pi}{M_P^2}\left(- \mathcal{F}\mathcal{H}^2\right),
\label{eq:tautauEE}
\ee 
where $k$ is the spacetime curvature. This ansatz satisfies the equations of motion for H, and from the Bianchi identity, $\epsilon^{\mu\nu\rho\varphi}\partial_\rho H_{\varphi\mu\nu}=0$, we can obtain
\be 
\partial_r(\mathcal{H}\beta^3\varepsilon_{ijk}) = 0, 
\ee 
where we have used $\epsilon_{ijk} = \sqrt{|g|}\varepsilon_{ijk}$. This necessitates that 
\be 
\mathcal{H} =\frac{\mathcal{H}_0}{\beta^3}, 
\label{eq:calHGS}
\ee 
where $\mathcal{H}_0$ is a dimensionless constant. This constant can be fixed by normalizing the integral of the field strength over the three sphere:
\be 
\int_{S_3}\mathcal{H}_0d\Omega_{3} = 2\pi^2\mathcal{H}_0 = n
\ee 
to obtain 
\be 
\mathcal{H}_0 = \frac{n}{2\pi^2}. 
\ee 
where n is an integer number that represents the charge of the instanton. Now, restricting attention to the three-sphere, let $\beta^2 = \tau^2$, and $k=1$ . Then, plugging in Eq.~\eqref{eq:calHGS} into Eq.~\eqref{eq:tautauEE}, we obtain:
\be 
\alpha^2 = \left(1 - \frac{n^2}{3\pi^3 M_P^2f_a^2 \tau^4}\right)^{-1} = \left(1 - \frac{L^4}{\tau^4}\right)^{-1}, 
\ee 
where $L$ is a length scale that describes the minimum radius of the wormhole throat. 

Thus, the metric for the Giddings-Strominger wormhole is given by 
\be 
ds^2 = \frac{1}{\sqrt{1-L^4/\tau^4}}d\tau^2 + \tau^2d\Omega_3^3.
\ee 
Then, from the trace of the Einstein equations, we can calculate the instanton action and find 
\be 
 S_{W} = \frac{n^2}{8\pi L^2 f_a^2}, 
\ee 
where we have performed an integral over $\tau$ and the three-sphere, $\Omega_3$. The quantity $L$ can also be written in terms of $f_a$ and $M_P$ as \cite{Alonso:2017avz}
\be 
L = \left(\frac{n^2}{3\pi^2 M_P^2 f_a^2}\right)^{1/4},
\ee 
to equivalently write this expression as 
\be 
S_{W}= \frac{\sqrt{3\pi}nM_P}{8f_a},
\ee 
which is the standard Giddings-Strominger wormhole result.

\bibliographystyle{unsrt}
\bibliography{master}

@article{PhysRevLett.96.081301,
  title = {Leptogenesis from Gravity Waves in Models of Inflation},
  author = {Alexander, Stephon H. S. and Peskin, Michael E. and Sheikh-Jabbari, M. M.},
  journal = {Phys. Rev. Lett.},
  volume = {96},
  issue = {8},
  pages = {081301},
  numpages = {4},
  year = {2006},
  month = {Feb},
  publisher = {American Physical Society},
  doi = {10.1103/PhysRevLett.96.081301},
  url = {https://link.aps.org/doi/10.1103/PhysRevLett.96.081301}
}

@article{Alexander:2014bsa,
      author         = "Alexander, Stephon and Cormack, Sam and Marcianò,
                        Antonino and Yunes, Nicolás",
      title          = "{Gravitational-Wave Mediated Preheating}",
      journal        = "Phys. Lett.",
      volume         = "B743",
      year           = "2015",
      pages          = "82-86",
      doi            = "10.1016/j.physletb.2015.02.018",
      eprint         = "1405.4288",
      archivePrefix  = "arXiv",
      primaryClass   = "gr-qc",
      reportNumber   = "NSF-KITP-14-047",
      SLACcitation   = "%%CITATION = ARXIV:1405.4288;%%"
}

@book{Polchinski:1998rr,
    author = "Polchinski, J.",
    title = "{String theory. Vol. 2: Superstring theory and beyond}",
    doi = "10.1017/CBO9780511618123",
    isbn = "978-0-511-25228-0, 978-0-521-63304-8, 978-0-521-67228-3",
    publisher = "Cambridge University Press",
    series = "Cambridge Monographs on Mathematical Physics",
    month = "12",
    year = "2007"
}

@article{Planck:2018vyg,
    author = "Aghanim, N. and others",
    collaboration = "Planck",
    title = "{Planck 2018 results. VI. Cosmological parameters}",
    eprint = "1807.06209",
    archivePrefix = "arXiv",
    primaryClass = "astro-ph.CO",
    doi = "10.1051/0004-6361/201833910",
    journal = "Astron. Astrophys.",
    volume = "641",
    pages = "A6",
    year = "2020",
    note = "[Erratum: Astron.Astrophys. 652, C4 (2021)]"
}

@article{Schafer:1996wv,
    author = {Sch{\"a}fer, Thomas and Shuryak, Edward V.},
    title = "{Instantons in QCD}",
    eprint = "hep-ph/9610451",
    archivePrefix = "arXiv",
    reportNumber = "DOE-ER-40561-293, INT-96-00-150",
    doi = "10.1103/RevModPhys.70.323",
    journal = "Rev. Mod. Phys.",
    volume = "70",
    pages = "323--426",
    year = "1998"
}

@Article{Smith:2007jm,
     author    = "Smith, Tristan L. and Erickcek, Adrienne L. and Caldwell,
                  Robert R. and Kamionkowski, Marc",
     title     = "{The effects of Chern-Simons gravity on bodies orbiting the
                  Earth}",
     journal   = "Phys. Rev.",
     volume    = "D77",
     year      = "2008",
     pages     = "024015",
     eprint    = "0708.0001",
     archivePrefix = "arXiv",
     primaryClass  =  "astro-ph",
     doi       = "10.1103/PhysRevD.77.024015",
     SLACcitation  = "%%CITATION = 0708.0001;%%"
}

@Article{Chung:1993ge,
    author = "Chung, J. and Hulbert, G.M.",
    title  = "A time integration algorithm for structural dynamics with improved numerical dissipation: the generalized-$\alpha$ method",
    journal = "Journal of Applied Mechanics, Transactions of the A.S.M.E.",
    volume  = "60",
    year = "1993",
    pages = "371-375"
}

@Article{Grumiller:2007rv,
     author    = "Grumiller, Daniel and Yunes, Nicolas",
     title     = "{How do Black Holes Spin in Chern-Simons Modified
                  Gravity?}",
     journal   = "Phys. Rev.",
     volume    = "D77",
     year      = "2008",
     pages     = "044015",
     eprint    = "0711.1868",
     archivePrefix = "arXiv",
     primaryClass  =  "gr-qc",
     doi       = "10.1103/PhysRevD.77.044015",
     SLACcitation  = "%%CITATION = 0711.1868;%%"
}

@article{Yunes:2010yf,
      author         = "Yunes, Nicolas and O'Shaughnessy, Richard and Owen,
                        Benjamin J. and Alexander, Stephon",
      title          = "{Testing gravitational parity violation with coincident
                        gravitational waves and short gamma-ray bursts}",
      journal        = "Phys.Rev.",
      volume         = "D82",
      pages          = "064017",
      doi            = "10.1103/PhysRevD.82.064017",
      year           = "2010",
      eprint         = "1005.3310",
      archivePrefix  = "arXiv",
      primaryClass   = "gr-qc",
}

@article{Silva:2020acr,
    author = "Silva, Hector O. and Holgado, A. Miguel and C{\'a}rdenas-Avenda{\~n}o, Alejandro and Yunes, Nicol{\'a}s",
    title = "{Astrophysical and theoretical physics implications from multimessenger neutron star observations}",
    eprint = "2004.01253",
    archivePrefix = "arXiv",
    primaryClass = "gr-qc",
    doi = "10.1103/PhysRevLett.126.181101",
    journal = "Phys. Rev. Lett.",
    volume = "126",
    number = "18",
    pages = "181101",
    year = "2021"
}

@article{Yagi:2012ya,
    author = "Yagi, Kent and Yunes, Nicolas and Tanaka, Takahiro",
    title = "{Slowly Rotating Black Holes in Dynamical Chern-Simons Gravity: Deformation Quadratic in the Spin}",
    eprint = "1206.6130",
    archivePrefix = "arXiv",
    primaryClass = "gr-qc",
    doi = "10.1103/PhysRevD.86.044037",
    journal = "Phys. Rev. D",
    volume = "86",
    pages = "044037",
    year = "2012",
    note = "[Erratum: Phys.Rev.D 89, 049902 (2014)]"
}

@article{Yunes:2009hc,
    author = "Yunes, Nicolas and Pretorius, Frans",
    title = "{Dynamical Chern-Simons Modified Gravity. I. Spinning Black Holes in the Slow-Rotation Approximation}",
    eprint = "0902.4669",
    archivePrefix = "arXiv",
    primaryClass = "gr-qc",
    doi = "10.1103/PhysRevD.79.084043",
    journal = "Phys. Rev. D",
    volume = "79",
    pages = "084043",
    year = "2009"
}

@book{Green:1987mn,
     author    = "Green, Michael B. and Schwarz, J. H. and Witten, Edward",
     title     = "Superstring Theory. Vol. 2: Loop Amplitides, Anomalies and
                  Phenomenology",
     publisher = "Cambridge University Press",
     address   = "Cambridge, UK",
     year     =  "1987"
}

@Article{Alexander:2004xd,
     author    = "Alexander, Stephon H. S. and Gates, S. James, Jr.",
     title     = "Can the string scale be related to the cosmic baryon
                  asymmetry?",
     journal   = "JCAP",
     volume    = "0606",
     year      = "2006",
     pages     = "018",
     eprint    = "hep-th/0409014",
     SLACcitation  = "%%CITATION = HEP-TH 0409014;%%"
}

@Article{Alexander:2007kv,
     author    = "Alexander, Stephon and Finn, Lee Samuel and Yunes, Nicolas
                  ",
     title     = "{A gravitational-wave probe of effective quantum gravity}",
     journal   = "Phys. Rev.",
     volume    = "D78",
     year      = "2008",
     pages     = "066005",
     eprint    = "0712.2542",
     archivePrefix = "arXiv",
     primaryClass  =  "gr-qc",
     doi       = "10.1103/PhysRevD.78.066005",
     SLACcitation  = "%%CITATION = 0712.2542;%%"
}

@ARTICLE{Alexander:2009tp,
   author = {{Alexander}, S. and {Yunes}, N.},
    title = "{Chern-Simons modified general relativity}",
  journal = "{Phys. Rep.}",
archivePrefix = "arXiv",
   eprint = {0907.2562},
 primaryClass = "hep-th",
     year = 2009,
    month = aug,
   volume = 480,
    pages = {1-55},
      doi = {10.1016/j.physrep.2009.07.002},
   adsurl = {http://adsabs.harvard.edu/abs/2009PhR...480....1A}
}

@Article{sopuertayunes,
     author    = "Sopuerta, Carlos F. and Yunes, Nicolas",
     title     = "{New Kludge Scheme for the Construction of Approximate
                  Waveforms for Extreme-Mass-Ratio Inspirals}",
     journal   = "Phys. Rev.",
     volume    = "D84",
     year      = "2011",
     pages     = "124060",
     eprint    = "1109.0572",
     archivePrefix = "arXiv",
     primaryClass  =  "gr-qc",
     doi       = "10.1103/PhysRevD.84.124060",
     SLACcitation  = "%%CITATION = 1109.0572;%%"
}

@Article{jackiw,
     author    = "Jackiw, R. and Pi, S. Y.",
     title     = "{Chern-Simons modification of general relativity}",
     journal   = "Phys. Rev.",
     volume    = "D68",
     year      = "2003",
     pages     = "104012",
     eprint    = "gr-qc/03x1",
     archivePrefix = "arXiv",
     doi       = "10.1103/PhysRevD.68.104012",
     SLACcitation  = "%%CITATION = GR-QC/0308071;%%"
}

@article{Alexander:2017jmt,
      author         = "Alexander, Stephon H. and Yunes, Nicolas",
      title          = "{Gravitational wave probes of parity violation in compact
                        binary coalescences}",
      journal        = "Phys. Rev.",
      volume         = "D97",
      year           = "2018",
      number         = "6",
      pages          = "064033",
      doi            = "10.1103/PhysRevD.97.064033",
      eprint         = "1712.01853",
      archivePrefix  = "arXiv",
      primaryClass   = "gr-qc",
      SLACcitation   = "%%CITATION = ARXIV:1712.01853;%%"
}

@article{ALVAREZGAUME1984269,
title = {Gravitational anomalies},
journal = {Nuclear Physics B},
volume = {234},
number = {2},
pages = {269-330},
year = {1984},
issn = {0550-3213},
doi = {https://doi.org/10.1016/0550-3213(84)90066-X},
author = {Luis Alvarez-Gaume and Edward Witten}
}

@article{Shiromizu:2013pna,
    author = "Shiromizu, Tetsuya and Tanabe, Kentaro",
    title = "{Static spacetimes with/without black holes in dynamical Chern-Simons gravity}",
    eprint = "1303.6056",
    archivePrefix = "arXiv",
    primaryClass = "gr-qc",
    doi = "10.1103/PhysRevD.87.081504",
    journal = "Phys. Rev. D",
    volume = "87",
    pages = "081504",
    year = "2013"
}

@article{Delbourgo:1972xb,
    author = "Delbourgo, Robert and Salam, Abdus",
    title = "{The gravitational correction to pcac}",
    reportNumber = "ICTP-71-24",
    doi = "10.1016/0370-2693(72)90825-8",
    journal = "Phys. Lett. B",
    volume = "40",
    pages = "381--382",
    year = "1972"
}

@article{Arvanitaki:2009fg,
    author = "Arvanitaki, Asimina and Dimopoulos, Savas and Dubovsky, Sergei and Kaloper, Nemanja and March-Russell, John",
    title = "{String Axiverse}",
    eprint = "0905.4720",
    archivePrefix = "arXiv",
    primaryClass = "hep-th",
    doi = "10.1103/PhysRevD.81.123530",
    journal = "Phys. Rev. D",
    volume = "81",
    pages = "123530",
    year = "2010"
}

@article{Ezquiaga:2017ekz,
    author = "Ezquiaga, Jose Mar\'\i{}a and Zumalac\'arregui, Miguel",
    title = "{Dark Energy After GW170817: Dead Ends and the Road Ahead}",
    eprint = "1710.05901",
    archivePrefix = "arXiv",
    primaryClass = "astro-ph.CO",
    reportNumber = "IFT-UAM-CSIC-17-096, NORDITA-2017-109",
    doi = "10.1103/PhysRevLett.119.251304",
    journal = "Phys. Rev. Lett.",
    volume = "119",
    number = "25",
    pages = "251304",
    year = "2017"
}

@article{Alexander:2022avt,
    author = "Alexander, Stephon and Gabadadze, Gregory and Jenks, Leah and Yunes, Nicol{\'a}s",
    title = "{Black hole superradiance in dynamical Chern-Simons gravity}",
    eprint = "2201.02220",
    archivePrefix = "arXiv",
    primaryClass = "gr-qc",
    doi = "10.1103/PhysRevD.107.084016",
    journal = "Phys. Rev. D",
    volume = "107",
    number = "8",
    pages = "084016",
    year = "2023"
}

@article{Lue:1998mq,
    author = "Lue, Arthur and Wang, Li-Min and Kamionkowski, Marc",
    title = "{Cosmological signature of new parity violating interactions}",
    eprint = "astro-ph/9812088",
    archivePrefix = "arXiv",
    reportNumber = "CU-TP-926, CAL-675",
    doi = "10.1103/PhysRevLett.83.1506",
    journal = "Phys. Rev. Lett.",
    volume = "83",
    pages = "1506--1509",
    year = "1999"
}

@article{Giddings:1987cg,
    author = "Giddings, Steven B. and Strominger, Andrew",
    title = "{Axion Induced Topology Change in Quantum Gravity and String Theory}",
    reportNumber = "HUTP-87-A067",
    doi = "10.1016/0550-3213(88)90446-4",
    journal = "Nucl. Phys. B",
    volume = "306",
    pages = "890--907",
    year = "1988"
}

@article{Alonso:2017avz,
    author = "Alonso, Rodrigo and Urbano, Alfredo",
    title = "{Wormholes and masses for Goldstone bosons}",
    eprint = "1706.07415",
    archivePrefix = "arXiv",
    primaryClass = "hep-ph",
    reportNumber = "CERN-TH-2017-135",
    doi = "10.1007/JHEP02(2019)136",
    journal = "JHEP",
    volume = "02",
    pages = "136",
    year = "2019"
}

@article{Hebecker:2018ofv,
    author = "Hebecker, Arthur and Mikhail, Thomas and Soler, Pablo",
    title = "{Euclidean wormholes, baby universes, and their impact on particle physics and cosmology}",
    eprint = "1807.00824",
    archivePrefix = "arXiv",
    primaryClass = "hep-th",
    doi = "10.3389/fspas.2018.00035",
    journal = "Front. Astron. Space Sci.",
    volume = "5",
    pages = "35",
    year = "2018"
}

@article{Lee:1988ge,
    author = "Lee, Ki-Myeong",
    title = "{Wormholes and Goldstone Bosons}",
    reportNumber = "FERMILAB-PUB-88-027-T",
    doi = "10.1103/PhysRevLett.61.263",
    journal = "Phys. Rev. Lett.",
    volume = "61",
    pages = "263--266",
    year = "1988"
}

@article{Alexander:2021ssr,
    author = "Alexander, Stephon and Gabadadze, Gregory and Jenks, Leah and Yunes, Nicol\'as",
    title = "{Chern-Simons caps for rotating black holes}",
    eprint = "2104.00019",
    archivePrefix = "arXiv",
    primaryClass = "hep-th",
    doi = "10.1103/PhysRevD.104.064033",
    journal = "Phys. Rev. D",
    volume = "104",
    number = "6",
    pages = "064033",
    year = "2021"
}

@article{Abbott:1989jw,
    author = "Abbott, L. F. and Wise, Mark B.",
    title = "{Wormholes and Global Symmetries}",
    reportNumber = "CALT-68-1523",
    doi = "10.1016/0550-3213(89)90503-8",
    journal = "Nucl. Phys. B",
    volume = "325",
    pages = "687--704",
    year = "1989"
}

@article{Hui:2021tkt,
    author = "Hui, Lam",
    title = "{Wave Dark Matter}",
    eprint = "2101.11735",
    archivePrefix = "arXiv",
    primaryClass = "astro-ph.CO",
    doi = "10.1146/annurev-astro-120920-010024",
    journal = "Ann. Rev. Astron. Astrophys.",
    volume = "59",
    pages = "247--289",
    year = "2021"
}

@article{Capanelli:2023uwv,
    author = "Capanelli, Christian and Jenks, Leah and Kolb, Edward W. and McDonough, Evan",
    title = "{Cosmological implications of Kalb-Ramond-like particles}",
    eprint = "2309.02485",
    archivePrefix = "arXiv",
    primaryClass = "hep-ph",
    doi = "10.1007/JHEP06(2024)075",
    journal = "JHEP",
    volume = "06",
    pages = "075",
    year = "2024"
}

@article{Alexander:2022cow,
    author = "Alexander, Stephon and Creque-Sarbinowski, Cyril",
    title = "{UV completions of Chern-Simons gravity}",
    eprint = "2207.05094",
    archivePrefix = "arXiv",
    primaryClass = "hep-ph",
    doi = "10.1103/PhysRevD.108.104046",
    journal = "Phys. Rev. D",
    volume = "108",
    number = "10",
    pages = "104046",
    year = "2023"
}

@article{Richards:2023xsr,
    author = "Richards, Chloe and Dima, Alexandru and Witek, Helvi",
    title = "{Black holes in massive dynamical Chern-Simons gravity: Scalar hair and quasibound states at decoupling}",
    eprint = "2305.07704",
    archivePrefix = "arXiv",
    primaryClass = "gr-qc",
    doi = "10.1103/PhysRevD.108.044078",
    journal = "Phys. Rev. D",
    volume = "108",
    number = "4",
    pages = "044078",
    year = "2023"
}

@article{Nashed:2022kes,
    author = "Nashed, G. G. L. and Nojiri, Shin'ichi",
    title = "{Slow-rotating black holes with potential in dynamical Chern-Simons modified gravitational theory}",
    eprint = "2208.11498",
    archivePrefix = "arXiv",
    primaryClass = "gr-qc",
    doi = "10.1088/1475-7516/2023/02/033",
    journal = "JCAP",
    volume = "02",
    pages = "033",
    year = "2023"
}

@article{Rey:1989mg,
    author = "Rey, Soo-Jong",
    title = "{The Axion Dynamics in Wormhole Background}",
    reportNumber = "PRINT-89-0095 (UC,SANTA-BARBARA)",
    doi = "10.1103/PhysRevD.39.3185",
    journal = "Phys. Rev. D",
    volume = "39",
    pages = "3185",
    year = "1989"
}

@article{Cardoso:2009pk,
    author = "Cardoso, Vitor and Gualtieri, Leonardo",
    title = "{Perturbations of Schwarzschild black holes in Dynamical Chern-Simons modified gravity}",
    eprint = "0907.5008",
    archivePrefix = "arXiv",
    primaryClass = "gr-qc",
    doi = "10.1103/PhysRevD.81.089903",
    journal = "Phys. Rev. D",
    volume = "80",
    pages = "064008",
    year = "2009",
    note = "[Erratum: Phys.Rev.D 81, 089903 (2010)]"
}

@article{Jenks:2023pmk,
    author = "Jenks, Leah and Choi, Lyla and Lagos, Macarena and Yunes, Nicol\'as",
    title = "{Parametrized parity violation in gravitational wave propagation}",
    eprint = "2305.10478",
    archivePrefix = "arXiv",
    primaryClass = "gr-qc",
    doi = "10.1103/PhysRevD.108.044023",
    journal = "Phys. Rev. D",
    volume = "108",
    number = "4",
    pages = "044023",
    year = "2023"
}

@article{Lagos:2024boe,
    author = "Lagos, Macarena and Jenks, Leah and Isi, Maximiliano and Hotokezaka, Kenta and Metzger, Brian D. and Burns, Eric and Farr, Will M. and Perkins, Scott and Wong, Kaze W. K. and Yunes, Nicolas",
    title = "{Birefringence tests of gravity with multimessenger binaries}",
    eprint = "2402.05316",
    archivePrefix = "arXiv",
    primaryClass = "gr-qc",
    doi = "10.1103/PhysRevD.109.124003",
    journal = "Phys. Rev. D",
    volume = "109",
    number = "12",
    pages = "124003",
    year = "2024"
}

@article{Daniel:2024lev,
    author = "Daniel, Tatsuya and Jenks, Leah and Alexander, Stephon",
    title = "{Gravitational waves in Chern-Simons-Gauss-Bonnet gravity}",
    eprint = "2403.09373",
    archivePrefix = "arXiv",
    primaryClass = "gr-qc",
    doi = "10.1103/PhysRevD.109.124012",
    journal = "Phys. Rev. D",
    volume = "109",
    number = "12",
    pages = "124012",
    year = "2024"
}

@article{Callister:2023tws,
    author = "Callister, Thomas and Jenks, Leah and Holz, Daniel E. and Yunes, Nicol{\'a}s",
    title = "{New probe of gravitational parity violation through nonobservation of the stochastic gravitational-wave background}",
    eprint = "2312.12532",
    archivePrefix = "arXiv",
    primaryClass = "gr-qc",
    doi = "10.1103/PhysRevD.111.044041",
    journal = "Phys. Rev. D",
    volume = "111",
    number = "4",
    pages = "044041",
    year = "2025"
}

@article{Inomata:2024ald,
    author = "Inomata, Keisuke and Jenks, Leah and Kamionkowski, Marc",
    title = "{Parity-breaking galaxy 4-point function from lensing by chiral gravitational waves}",
    eprint = "2408.03994",
    archivePrefix = "arXiv",
    primaryClass = "astro-ph.CO",
    doi = "10.1103/PhysRevD.111.043504",
    journal = "Phys. Rev. D",
    volume = "111",
    number = "4",
    pages = "043504",
    year = "2025"
}

@article{Alexander:lepto,
    author = "Alexander, Stephon Haigh-Solom and Peskin, Michael E. and Sheikh-Jabbari, Mohammad M.",
    title = "{Leptogenesis from gravity waves in models of inflation}",
    eprint = "hep-th/0403069",
    archivePrefix = "arXiv",
    reportNumber = "SLAC-PUB-10226, SU-ITP-04-08",
    doi = "10.1103/PhysRevLett.96.081301",
    journal = "Phys. Rev. Lett.",
    volume = "96",
    pages = "081301",
    year = "2006"
}

@article{Alexander:2011hz,
    author = "Alexander, Stephon and Marciano, Antonino and Spergel, David",
    title = "{Chern-Simons Inflation and Baryogenesis}",
    eprint = "1107.0318",
    archivePrefix = "arXiv",
    primaryClass = "hep-th",
    doi = "10.1088/1475-7516/2013/04/046",
    journal = "JCAP",
    volume = "04",
    pages = "046",
    year = "2013"
}

@article{Cheong:2024kum,
    author = "Cheong, Dhong Yeon and Hamaguchi, Koichi and Kanazawa, Yoshiki and Lee, Sung Mook and Nagata, Natsumi and Park, Seong Chan",
    title = "{Wormhole-induced ALP dark matter}",
    eprint = "2411.07713",
    archivePrefix = "arXiv",
    primaryClass = "hep-ph",
    reportNumber = "CERN-TH-2024-195",
    doi = "10.1007/JHEP02(2025)183",
    journal = "JHEP",
    volume = "02",
    pages = "183",
    year = "2025"
}

@article{Cheong:2023hrj,
    author = "Cheong, Dhong Yeon and Park, Seong Chan and Shin, Chang Sub",
    title = "{Effective theory approach for axion wormholes}",
    eprint = "2310.11260",
    archivePrefix = "arXiv",
    primaryClass = "hep-th",
    reportNumber = "CERN-TH-2023-184",
    doi = "10.1007/JHEP07(2024)039",
    journal = "JHEP",
    volume = "07",
    pages = "039",
    year = "2024"
}

@article{Klebanov:1988eh,
    author = "Klebanov, Igor R. and Susskind, Leonard and Banks, Tom",
    title = "{Wormholes and the Cosmological Constant}",
    reportNumber = "SLAC-PUB-4705",
    doi = "10.1016/0550-3213(89)90538-5",
    journal = "Nucl. Phys. B",
    volume = "317",
    pages = "665--692",
    year = "1989"
}

@article{Tsujikawa:2013fta,
    author = "Tsujikawa, Shinji",
    title = "{Quintessence: A Review}",
    eprint = "1304.1961",
    archivePrefix = "arXiv",
    primaryClass = "gr-qc",
    doi = "10.1088/0264-9381/30/21/214003",
    journal = "Class. Quant. Grav.",
    volume = "30",
    pages = "214003",
    year = "2013"
}

@article{Koivisto:2006ai,
    author = "Koivisto, Tomi and Mota, David F.",
    title = "{Gauss-Bonnet Quintessence: Background Evolution, Large Scale Structure and Cosmological Constraints}",
    eprint = "hep-th/0609155",
    archivePrefix = "arXiv",
    doi = "10.1103/PhysRevD.75.023518",
    journal = "Phys. Rev. D",
    volume = "75",
    pages = "023518",
    year = "2007"
}

@article{Shajib:2025tpd,
    author = "Shajib, Anowar J. and Frieman, Joshua A.",
    title = "{Scalar-field dark energy models: Current and forecast constraints}",
    eprint = "2502.06929",
    archivePrefix = "arXiv",
    primaryClass = "astro-ph.CO",
    doi = "10.1103/kjpb-r698",
    journal = "Phys. Rev. D",
    volume = "112",
    number = "6",
    pages = "063508",
    year = "2025"
}

@article{DESI:2025fii,
    author = "Lodha, K. and others",
    collaboration = "DESI",
    title = "{Extended dark energy analysis using DESI DR2 BAO measurements}",
    eprint = "2503.14743",
    archivePrefix = "arXiv",
    primaryClass = "astro-ph.CO",
    reportNumber = "FERMILAB-PUB-25-0164-PPD",
    doi = "10.1103/w4c6-1r5j",
    journal = "Phys. Rev. D",
    volume = "112",
    number = "8",
    pages = "083511",
    year = "2025"
}

@article{DESI:2025zgx,
    author = "Abdul Karim, M. and others",
    collaboration = "DESI",
    title = "{DESI DR2 results. II. Measurements of baryon acoustic oscillations and cosmological constraints}",
    eprint = "2503.14738",
    archivePrefix = "arXiv",
    primaryClass = "astro-ph.CO",
    reportNumber = "FERMILAB-PUB-25-0169-PPD",
    doi = "10.1103/tr6y-kpc6",
    journal = "Phys. Rev. D",
    volume = "112",
    number = "8",
    pages = "083515",
    year = "2025"
}

@article{Frieman:1995pm,
    author = "Frieman, Joshua A. and Hill, Christopher T. and Stebbins, Albert and Waga, Ioav",
    title = "{Cosmology with ultralight pseudo Nambu-Goldstone bosons}",
    eprint = "astro-ph/9505060",
    archivePrefix = "arXiv",
    reportNumber = "FERMILAB-PUB-95-066-A",
    doi = "10.1103/PhysRevLett.75.2077",
    journal = "Phys. Rev. Lett.",
    volume = "75",
    pages = "2077--2080",
    year = "1995"
}

@article{Nomura:2000yk,
    author = "Nomura, Yasunori and Watari, T. and Yanagida, T.",
    title = "{Quintessence axion potential induced by electroweak instanton effects}",
    eprint = "hep-ph/0004182",
    archivePrefix = "arXiv",
    reportNumber = "UT-883",
    doi = "10.1016/S0370-2693(00)00605-5",
    journal = "Phys. Lett. B",
    volume = "484",
    pages = "103--111",
    year = "2000"
}

@article{Peccei:1977hh,
    author = "Peccei, R. D. and Quinn, Helen R.",
    title = "{CP Conservation in the Presence of Instantons}",
    reportNumber = "ITP-568-STANFORD",
    doi = "10.1103/PhysRevLett.38.1440",
    journal = "Phys. Rev. Lett.",
    volume = "38",
    pages = "1440--1443",
    year = "1977"
}

@article{Peccei:1977ur,
    author = "Peccei, R. D. and Quinn, Helen R.",
    title = "{Constraints Imposed by CP Conservation in the Presence of Instantons}",
    reportNumber = "ITP-572-STANFORD",
    doi = "10.1103/PhysRevD.16.1791",
    journal = "Phys. Rev. D",
    volume = "16",
    pages = "1791--1797",
    year = "1977"
}

@article{Wilczek:1977pj,
    author = "Wilczek, Frank",
    title = "{Problem of Strong  $P$  and  $T$  Invariance in the Presence of Instantons}",
    reportNumber = "Print-77-0939 (COLUMBIA)",
    doi = "10.1103/PhysRevLett.40.279",
    journal = "Phys. Rev. Lett.",
    volume = "40",
    pages = "279--282",
    year = "1978"
}

@article{Weinberg:1977ma,
    author = "Weinberg, Steven",
    title = "{A New Light Boson?}",
    reportNumber = "HUTP-77/A074",
    doi = "10.1103/PhysRevLett.40.223",
    journal = "Phys. Rev. Lett.",
    volume = "40",
    pages = "223--226",
    year = "1978"
}

@article{Hu:2000ke,
    author = "Hu, Wayne and Barkana, Rennan and Gruzinov, Andrei",
    title = "{Cold and fuzzy dark matter}",
    eprint = "astro-ph/0003365",
    archivePrefix = "arXiv",
    doi = "10.1103/PhysRevLett.85.1158",
    journal = "Phys. Rev. Lett.",
    volume = "85",
    pages = "1158--1161",
    year = "2000"
}

@article{Kolb:2023ydq,
    author = "Kolb, Edward W. and Long, Andrew J.",
    title = "{Cosmological gravitational particle production and its implications for cosmological relics}",
    eprint = "2312.09042",
    archivePrefix = "arXiv",
    primaryClass = "astro-ph.CO",
    doi = "10.1103/RevModPhys.96.045005",
    journal = "Rev. Mod. Phys.",
    volume = "96",
    number = "4",
    pages = "045005",
    year = "2024"
}

@article{Svrcek:2006yi,
    author = "Svrcek, Peter and Witten, Edward",
    title = "{Axions In String Theory}",
    eprint = "hep-th/0605206",
    archivePrefix = "arXiv",
    reportNumber = "SLAC-PUB-11894",
    doi = "10.1088/1126-6708/2006/06/051",
    journal = "JHEP",
    volume = "06",
    pages = "051",
    year = "2006"
}

@article{Masso:1995tw,
    author = "Masso, Eduard and Toldra, Ramon",
    title = "{On a light spinless particle coupled to photons}",
    eprint = "hep-ph/9503293",
    archivePrefix = "arXiv",
    reportNumber = "UAB-FT-361",
    doi = "10.1103/PhysRevD.52.1755",
    journal = "Phys. Rev. D",
    volume = "52",
    pages = "1755--1763",
    year = "1995"
}

@article{Masso:2002ip,
    author = "Masso, Eduard",
    editor = "Morales, A. and Morales, J. and Cebrian, S.",
    title = "{Axions and axion like particles}",
    eprint = "hep-ph/0209132",
    archivePrefix = "arXiv",
    reportNumber = "UAB-FT-530",
    doi = "10.1016/S0920-5632(02)01893-5",
    journal = "Nucl. Phys. B Proc. Suppl.",
    volume = "114",
    pages = "67--73",
    year = "2003"
}

@article{Conlon:2006tq,
    author = "Conlon, Joseph P.",
    title = "{The QCD axion and moduli stabilisation}",
    eprint = "hep-th/0602233",
    archivePrefix = "arXiv",
    reportNumber = "DAMTP-2006-17",
    doi = "10.1088/1126-6708/2006/05/078",
    journal = "JHEP",
    volume = "05",
    pages = "078",
    year = "2006"
}

@article{Marsh:2015xka,
    author = "Marsh, David J. E.",
    title = "{Axion Cosmology}",
    eprint = "1510.07633",
    archivePrefix = "arXiv",
    primaryClass = "astro-ph.CO",
    reportNumber = "KCL-PH-TH-2015-50",
    doi = "10.1016/j.physrep.2016.06.005",
    journal = "Phys. Rept.",
    volume = "643",
    pages = "1--79",
    year = "2016"
}

@article{Sikivie:2006ni,
    author = "Sikivie, Pierre",
    editor = "Kuster, Markus and Raffelt, Georg and Beltran, Berta",
    title = "{Axion Cosmology}",
    eprint = "astro-ph/0610440",
    archivePrefix = "arXiv",
    reportNumber = "UFIFT-HEP-06-16",
    doi = "10.1007/978-3-540-73518-2_2",
    journal = "Lect. Notes Phys.",
    volume = "741",
    pages = "19--50",
    year = "2008"
}

@article{Coleman:1988tj,
    author = "Coleman, Sidney R.",
    title = "{Why There Is Nothing Rather Than Something: A Theory of the Cosmological Constant}",
    reportNumber = "HUTP-88/A022",
    doi = "10.1016/0550-3213(88)90097-1",
    journal = "Nucl. Phys. B",
    volume = "310",
    pages = "643--668",
    year = "1988"
}

@article{Preskill:1988na,
    author = "Preskill, John",
    title = "{Wormholes in Space-time and the Constants of Nature}",
    reportNumber = "CALT-68-1521",
    doi = "10.1016/0550-3213(89)90592-0",
    journal = "Nucl. Phys. B",
    volume = "323",
    pages = "141--186",
    year = "1989"
}

@article{Grinstein:1988wr,
    author = "Grinstein, Benjamin and Wise, Mark B.",
    title = "{Light Scalars in Quantum Gravity}",
    reportNumber = "CALT-68-1505",
    doi = "10.1016/0370-2693(88)91788-1",
    journal = "Phys. Lett. B",
    volume = "212",
    pages = "407--410",
    year = "1988"
}

@article{Choi:1989ck,
    author = "Choi, Kiwoon and Holman, R.",
    title = "{A Wormhole Solution to the Strong {CP} Problem}",
    reportNumber = "CMU-HEP89-04",
    doi = "10.1103/PhysRevLett.62.2575",
    journal = "Phys. Rev. Lett.",
    volume = "62",
    pages = "2575",
    year = "1989"
}

@article{Preskill:1989zu,
    author = "Preskill, John and Trivedi, Sandip P. and Wise, Mark B.",
    title = "{Wormholes in Space-time and $\theta$ ({QCD})}",
    reportNumber = "CALT-68-1539",
    doi = "10.1016/0370-2693(89)90913-1",
    journal = "Phys. Lett. B",
    volume = "223",
    pages = "26--31",
    year = "1989"
}

@article{Brown:1989df,
    author = "Brown, J. David and Burgess, C. P. and Kshirsagar, A. and Whiting, Bernard F. and York, Jr., James W.",
    title = "{SCALAR FIELD WORMHOLES}",
    reportNumber = "IFP-340-UNC",
    doi = "10.1016/0550-3213(89)90101-6",
    journal = "Nucl. Phys. B",
    volume = "328",
    pages = "213--222",
    year = "1989"
}

@article{Kallosh:1995hi,
    author = "Kallosh, Renata and Linde, Andrei D. and Linde, Dmitri A. and Susskind, Leonard",
    title = "{Gravity and global symmetries}",
    eprint = "hep-th/9502069",
    archivePrefix = "arXiv",
    reportNumber = "SU-ITP-95-2",
    doi = "10.1103/PhysRevD.52.912",
    journal = "Phys. Rev. D",
    volume = "52",
    pages = "912--935",
    year = "1995"
}

@article{DESI:2024mwx,
    author = "Adame, A. G. and others",
    collaboration = "DESI",
    title = "{DESI 2024 VI: cosmological constraints from the measurements of baryon acoustic oscillations}",
    eprint = "2404.03002",
    archivePrefix = "arXiv",
    primaryClass = "astro-ph.CO",
    reportNumber = "FERMILAB-PUB-24-0154-PPD",
    doi = "10.1088/1475-7516/2025/02/021",
    journal = "JCAP",
    volume = "02",
    pages = "021",
    year = "2025"
}

@book{Kolb:1990vq,
    author = "Kolb, Edward W. and Turner, Michael S.",
    title = "{The Early Universe}",
    reportNumber = "FERMILAB-BOOK-1990-01",
    doi = "10.1201/9780429492860",
    isbn = "978-0-429-49286-0, 978-0-201-62674-2",
    publisher = "Taylor and Francis",
    volume = "69",
    month = "5",
    year = "2019"
}

@article{OHare:2024nmr,
    author = "O'Hare, Ciaran A. J.",
    title = "{Cosmology of axion dark matter}",
    eprint = "2403.17697",
    archivePrefix = "arXiv",
    primaryClass = "hep-ph",
    doi = "10.22323/1.454.0040",
    journal = "PoS",
    volume = "COSMICWISPers",
    pages = "040",
    year = "2024"
}

@article{Vilenkin:1984ib,
    author = "Vilenkin, Alexander",
    title = "{Cosmic Strings and Domain Walls}",
    reportNumber = "PRINT-84-0840 (TUFTS)",
    doi = "10.1016/0370-1573(85)90033-X",
    journal = "Phys. Rept.",
    volume = "121",
    pages = "263--315",
    year = "1985"
}

@article{Parker:1969au,
    author = "Parker, Leonard",
    title = "{Quantized fields and particle creation in expanding universes. 1.}",
    doi = "10.1103/PhysRev.183.1057",
    journal = "Phys. Rev.",
    volume = "183",
    pages = "1057--1068",
    year = "1969"
}

@article{Parker:1971pt,
    author = "Parker, L.",
    title = "{Quantized fields and particle creation in expanding universes. 2.}",
    doi = "10.1103/PhysRevD.3.346",
    journal = "Phys. Rev. D",
    volume = "3",
    pages = "346--356",
    year = "1971",
    note = "[Erratum: Phys.Rev.D 3, 2546--2546 (1971)]"
}

@article{Ford:2021syk,
    author = "Ford, L. H.",
    title = "{Cosmological particle production: a review}",
    eprint = "2112.02444",
    archivePrefix = "arXiv",
    primaryClass = "gr-qc",
    doi = "10.1088/1361-6633/ac1b23",
    journal = "Rept. Prog. Phys.",
    volume = "84",
    number = "11",
    year = "2021"
}

@article{Chung:1998zb,
    author = "Chung, Daniel J. H. and Kolb, Edward W. and Riotto, Antonio",
    title = "{Superheavy dark matter}",
    eprint = "hep-ph/9802238",
    archivePrefix = "arXiv",
    reportNumber = "FERMILAB-PUB-98-021-A, CERN-TH-98-37, OUTP-98-02-P, FERMILAB-FERMILAB-PUB-98-021-A, CERN-CERN-TH-98-37, OXFORD-U. --OUTP-98-02-P",
    doi = "10.1103/PhysRevD.59.023501",
    journal = "Phys. Rev. D",
    volume = "59",
    pages = "023501",
    year = "1998"
}

@article{Chung:1998ua,
    author = "Chung, Daniel J. H. and Kolb, Edward W. and Riotto, Antonio",
    title = "{Nonthermal supermassive dark matter}",
    eprint = "hep-ph/9805473",
    archivePrefix = "arXiv",
    reportNumber = "FERMILAB-PUB-98-154-A, CERN-TH-98-160, OUTP-98-40-P",
    doi = "10.1103/PhysRevLett.81.4048",
    journal = "Phys. Rev. Lett.",
    volume = "81",
    pages = "4048--4051",
    year = "1998"
}

@article{Graham:2015rva,
    author = "Graham, Peter W. and Mardon, Jeremy and Rajendran, Surjeet",
    title = "{Vector Dark Matter from Inflationary Fluctuations}",
    eprint = "1504.02102",
    archivePrefix = "arXiv",
    primaryClass = "hep-ph",
    doi = "10.1103/PhysRevD.93.103520",
    journal = "Phys. Rev. D",
    volume = "93",
    number = "10",
    pages = "103520",
    year = "2016"
}

@article{Ema:2019yrd,
    author = "Ema, Yohei and Nakayama, Kazunori and Tang, Yong",
    title = "{Production of purely gravitational dark matter: the case of fermion and vector boson}",
    eprint = "1903.10973",
    archivePrefix = "arXiv",
    primaryClass = "hep-ph",
    reportNumber = "DESY-19-050, DESY 19-050, KEK-TH-2114, UT-19-04",
    doi = "10.1007/JHEP07(2019)060",
    journal = "JHEP",
    volume = "07",
    pages = "060",
    year = "2019"
}

@article{Ema:2016hlw,
    author = "Ema, Yohei and Jinno, Ryusuke and Mukaida, Kyohei and Nakayama, Kazunori",
    title = "{Gravitational particle production in oscillating backgrounds and its cosmological implications}",
    eprint = "1604.08898",
    archivePrefix = "arXiv",
    primaryClass = "hep-ph",
    reportNumber = "UT-16-20, KEK-TH-1898, IPMU-16-0062",
    doi = "10.1103/PhysRevD.94.063517",
    journal = "Phys. Rev. D",
    volume = "94",
    number = "6",
    pages = "063517",
    year = "2016"
}

@article{Ema:2015dka,
    author = "Ema, Yohei and Jinno, Ryusuke and Mukaida, Kyohei and Nakayama, Kazunori",
    title = "{Gravitational Effects on Inflaton Decay}",
    eprint = "1502.02475",
    archivePrefix = "arXiv",
    primaryClass = "hep-ph",
    reportNumber = "UT-15-03",
    doi = "10.1088/1475-7516/2015/05/038",
    journal = "JCAP",
    volume = "05",
    pages = "038",
    year = "2015"
}

@article{Kolb:2020fwh,
    author = "Kolb, Edward W. and Long, Andrew J.",
    title = "{Completely dark photons from gravitational particle production during the inflationary era}",
    eprint = "2009.03828",
    archivePrefix = "arXiv",
    primaryClass = "astro-ph.CO",
    doi = "10.1007/JHEP03(2021)283",
    journal = "JHEP",
    volume = "03",
    pages = "283",
    year = "2021"
}

@article{SCHRODINGER1939899,
title = {The proper vibrations of the expanding universe},
journal = {Physica},
volume = {6},
number = {7},
pages = {899-912},
year = {1939},
issn = {0031-8914},
doi = {https://doi.org/10.1016/S0031-8914(39)90091-1},
url = {https://www.sciencedirect.com/science/article/pii/S0031891439900911},
author = {Erwin Schro¨dinger}
}

@article{Ahmed:2020fhc,
	Archiveprefix = {arXiv},
	Author = {Ahmed, Aqeel and Grzadkowski, Bohdan and Socha, Anna},
	Doi = {10.1007/JHEP08(2020)059},
	Eprint = {2005.01766},
	Journal = {JHEP},
	Pages = {059},
	Primaryclass = {hep-ph},
	Title = {{Gravitational production of vector dark matter}},
	Volume = {08},
	Year = {2020}}

@article{Alexander:2020gmv,
    author = "Alexander, Stephon and Jenks, Leah and McDonough, Evan",
    title = "{Higher spin dark matter}",
    eprint = "2010.15125",
    archivePrefix = "arXiv",
    primaryClass = "hep-ph",
    doi = "10.1016/j.physletb.2021.136436",
    journal = "Phys. Lett. B",
    volume = "819",
    pages = "136436",
    year = "2021"
}

@article{Kolb:2021nob,
    author = "Kolb, Edward W. and Long, Andrew J. and McDonough, Evan",
    title = "{Gravitino Swampland Conjecture}",
    eprint = "2103.10437",
    archivePrefix = "arXiv",
    primaryClass = "hep-th",
    doi = "10.1103/PhysRevLett.127.131603",
    journal = "Phys. Rev. Lett.",
    volume = "127",
    number = "13",
    pages = "131603",
    year = "2021"
}

@article{Kolb:2021xfn,
    author = "Kolb, Edward W. and Long, Andrew J. and McDonough, Evan",
    title = "{Catastrophic production of slow gravitinos}",
    eprint = "2102.10113",
    archivePrefix = "arXiv",
    primaryClass = "hep-th",
    doi = "10.1103/PhysRevD.104.075015",
    journal = "Phys. Rev. D",
    volume = "104",
    number = "7",
    pages = "075015",
    year = "2021"
}

@article{Maleknejad:2022gyf,
    author = "Maleknejad, Azadeh and McDonough, Evan",
    title = "{Ultralight pion and superheavy baryon dark matter}",
    eprint = "2205.12983",
    archivePrefix = "arXiv",
    primaryClass = "hep-ph",
    reportNumber = "CERN-TH-2022-086",
    doi = "10.1103/PhysRevD.106.095011",
    journal = "Phys. Rev. D",
    volume = "106",
    number = "9",
    pages = "095011",
    year = "2022"
}

@article{Jenks:2024fiu,
    author = "Jenks, Leah and Kolb, Edward W. and Thyme, Keyer",
    title = "{Gravitational particle production of scalars: analytic and numerical approaches including early reheating}",
    eprint = "2410.03938",
    archivePrefix = "arXiv",
    primaryClass = "hep-ph",
    doi = "10.1007/JHEP05(2025)077",
    journal = "JHEP",
    volume = "05",
    pages = "077",
    year = "2025"
}

@article{Capanelli:2024rlk,
    author = "Capanelli, Christian and Jenks, Leah and Kolb, Edward W. and McDonough, Evan",
    title = "{Gravitational production of completely dark photons with nonminimal couplings to gravity}",
    eprint = "2405.19390",
    archivePrefix = "arXiv",
    primaryClass = "hep-th",
    doi = "10.1007/JHEP09(2024)071",
    journal = "JHEP",
    volume = "09",
    pages = "071",
    year = "2024"
}

@article{Capanelli:2024pzd,
    author = "Capanelli, Christian and Jenks, Leah and Kolb, Edward W. and McDonough, Evan",
    title = "{Runaway Gravitational Production of Dark Photons}",
    eprint = "2403.15536",
    archivePrefix = "arXiv",
    primaryClass = "hep-th",
    doi = "10.1103/PhysRevLett.133.061602",
    journal = "Phys. Rev. Lett.",
    volume = "133",
    number = "6",
    pages = "061602",
    year = "2024"
}

@article{Kolb:2023dzp,
    author = "Kolb, Edward W. and Ling, Siyang and Long, Andrew J. and Rosen, Rachel A.",
    title = "{Cosmological gravitational particle production of massive spin-2 particles}",
    eprint = "2302.04390",
    archivePrefix = "arXiv",
    primaryClass = "astro-ph.CO",
    doi = "10.1007/JHEP05(2023)181",
    journal = "JHEP",
    volume = "05",
    pages = "181",
    year = "2023"
}

@article{Kaneta:2023uwi,
    author = "Kaneta, Kunio and Ke, Wenqi and Mambrini, Yann and Olive, Keith A. and Verner, Sarunas",
    title = "{Gravitational production of spin-3/2 particles during reheating}",
    eprint = "2309.15146",
    archivePrefix = "arXiv",
    primaryClass = "hep-ph",
    reportNumber = "UMN--TH--4225/23, FTPI--MINN--23/17, OU--HET--1204",
    doi = "10.1103/PhysRevD.108.115027",
    journal = "Phys. Rev. D",
    volume = "108",
    number = "11",
    pages = "115027",
    year = "2023"
}

@article{Racco:2024aac,
    author = "Racco, Davide and Verner, Sarunas and Xue, Wei",
    title = "{Gravitational production of heavy particles during and after inflation}",
    eprint = "2405.13883",
    archivePrefix = "arXiv",
    primaryClass = "hep-ph",
    doi = "10.1007/JHEP09(2024)129",
    journal = "JHEP",
    volume = "09",
    pages = "129",
    year = "2024"
}

@article{Chiba:2009sj,
    author = "Chiba, Takeshi",
    title = "{Slow-Roll Thawing Quintessence}",
    eprint = "0902.4037",
    archivePrefix = "arXiv",
    primaryClass = "astro-ph.CO",
    doi = "10.1103/PhysRevD.80.109902",
    journal = "Phys. Rev. D",
    volume = "79",
    pages = "083517",
    year = "2009",
    note = "[Erratum: Phys.Rev.D 80, 109902 (2009)]"
}

@article{Dutta:2008qn,
    author = "Dutta, Sourish and Scherrer, Robert J.",
    title = "{Hilltop Quintessence}",
    eprint = "0809.4441",
    archivePrefix = "arXiv",
    primaryClass = "astro-ph",
    doi = "10.1103/PhysRevD.78.123525",
    journal = "Phys. Rev. D",
    volume = "78",
    pages = "123525",
    year = "2008"
}

@article{Linder:2007wa,
    author = "Linder, Eric V.",
    title = "{The Dynamics of Quintessence, The Quintessence of Dynamics}",
    eprint = "0704.2064",
    archivePrefix = "arXiv",
    primaryClass = "astro-ph",
    doi = "10.1007/s10714-007-0550-z",
    journal = "Gen. Rel. Grav.",
    volume = "40",
    pages = "329--356",
    year = "2008"
}

@article{Carroll:1998zi,
    author = "Carroll, Sean M.",
    title = "{Quintessence and the rest of the world}",
    eprint = "astro-ph/9806099",
    archivePrefix = "arXiv",
    reportNumber = "NSF-ITP-98-063",
    doi = "10.1103/PhysRevLett.81.3067",
    journal = "Phys. Rev. Lett.",
    volume = "81",
    pages = "3067--3070",
    year = "1998"
}

@article{Figliolia:2025dtw,
    author = "Figliolia, Marco and Grippa, Francesco and Lambiase, Gaetano and Visinelli, Luca",
    title = "{Gravitational Signatures of Axion Dark Matter via Parity-Violating Interactions}",
    eprint = "2509.12038",
    archivePrefix = "arXiv",
    primaryClass = "astro-ph.CO",
    reportNumber = "CA21106; CA21136",
    month = "9",
    year = "2025"
}

@article{Ema:2021fdz,
    author = "Ema, Yohei and Mukaida, Kyohei and Nakayama, Kazunori",
    title = "{Scalar field couplings to quadratic curvature and decay into gravitons}",
    eprint = "2112.12774",
    archivePrefix = "arXiv",
    primaryClass = "hep-ph",
    reportNumber = "TU-1140, KEK-TH-2379",
    doi = "10.1007/JHEP05(2022)087",
    journal = "JHEP",
    volume = "05",
    pages = "087",
    year = "2022"
}

@article{Inomata:2018rin,
    author = "Inomata, Keisuke and Kamionkowski, Marc",
    title = "{Chiral photons from chiral gravitational waves}",
    eprint = "1811.04959",
    archivePrefix = "arXiv",
    primaryClass = "astro-ph.CO",
    reportNumber = "IPMU 18-0185",
    doi = "10.1103/PhysRevLett.123.031305",
    journal = "Phys. Rev. Lett.",
    volume = "123",
    number = "3",
    pages = "031305",
    year = "2019"
}

@article{Inomata:2018vbu,
    author = "Inomata, Keisuke and Kamionkowski, Marc",
    title = "{Circular polarization of the cosmic microwave background from vector and tensor perturbations}",
    eprint = "1811.04957",
    archivePrefix = "arXiv",
    primaryClass = "astro-ph.CO",
    reportNumber = "IPMU 18-0184",
    doi = "10.1103/PhysRevD.99.043501",
    journal = "Phys. Rev. D",
    volume = "99",
    number = "4",
    pages = "043501",
    year = "2019"
}

@article{Chowdhury:2025mye,
    author = "Chowdhury, Tammi and Jenks, Leah and Kolb, Edward W. and McDonough, Evan",
    title = "{Higgs Inflation: Particle Factory}",
    eprint = "2510.24651",
    archivePrefix = "arXiv",
    primaryClass = "hep-ph",
    month = "10",
    year = "2025"
}

@article{Berlin:2021txa,
    author = {Berlin, Asher and Blas, Diego and Tito D'Agnolo, Raffaele and Ellis, Sebastian A. R. and Harnik, Roni and Kahn, Yonatan and Sch{\"u}tte-Engel, Jan},
    title = "{Detecting high-frequency gravitational waves with microwave cavities}",
    eprint = "2112.11465",
    archivePrefix = "arXiv",
    primaryClass = "hep-ph",
    reportNumber = "FERMILAB-PUB-21-724-SQMS-T",
    doi = "10.1103/PhysRevD.105.116011",
    journal = "Phys. Rev. D",
    volume = "105",
    number = "11",
    pages = "116011",
    year = "2022"
}

@article{Kim:1979if,
    author = "Kim, Jihn E.",
    title = "{Weak Interaction Singlet and Strong CP Invariance}",
    reportNumber = "UPR-0120T",
    doi = "10.1103/PhysRevLett.43.103",
    journal = "Phys. Rev. Lett.",
    volume = "43",
    pages = "103",
    year = "1979"
}

@article{Shifman:1979if,
    author = "Shifman, Mikhail A. and Vainshtein, A. I. and Zakharov, Valentin I.",
    title = "{Can Confinement Ensure Natural CP Invariance of Strong Interactions?}",
    reportNumber = "ITEP-64-1979",
    doi = "10.1016/0550-3213(80)90209-6",
    journal = "Nucl. Phys. B",
    volume = "166",
    pages = "493--506",
    year = "1980"
}

@article{Zhitnitsky:1980tq,
    author = "Zhitnitsky, A. R.",
    title = "{On Possible Suppression of the Axion Hadron Interactions. (In Russian)}",
    journal = "Sov. J. Nucl. Phys.",
    volume = "31",
    pages = "260",
    year = "1980"
}

@article{Dine:1981rt,
    author = "Dine, Michael and Fischler, Willy and Srednicki, Mark",
    title = "{A Simple Solution to the Strong CP Problem with a Harmless Axion}",
    reportNumber = "Print-81-0320 (IAS,PRINCETON)",
    doi = "10.1016/0370-2693(81)90590-6",
    journal = "Phys. Lett. B",
    volume = "104",
    pages = "199--202",
    year = "1981"
}

\end{document}